\documentclass[longauth]{aa}

\usepackage{graphicx}
\usepackage{natbib}
\usepackage{scalerel}
\usepackage{booktabs}
\usepackage[table]{xcolor}
\newcommand*{\rt}{\textcolor{black}}

\usepackage{txfonts}
\usepackage[pdfencoding=auto,psdextra]{hyperref}
\hypersetup{
    colorlinks=true,
    linkcolor=blue,
    filecolor=magenta,      
    urlcolor=blue,
    citecolor=blue
}
\makeatletter
\renewcommand*\aa@pageof{, page \thepage{} of \pageref*{LastPage}}
\makeatother

\usepackage[utf8]{inputenc}

\usepackage[switch, modulo]{lineno}
 \usepackage{textcomp}

\usepackage{euclid}
\usepackage{orcidlink}

\begin{document}
%
%

\title{Euclid Quick Data Release (Q1)} \subtitle{The Strong Lensing Discovery Engine E -- Ensemble classification of strong gravitational lenses: lessons for Data Release 1}

\newcommand{\orcid}[1]{\orcidlink{#1}}

\author{Euclid Collaboration: P.~Holloway\orcid{0009-0002-8896-6100}\thanks{\email{philip.holloway@physics.ox.ac.uk}}\inst{\ref{aff1}}
\and A.~Verma\orcid{0000-0002-0730-0781}\inst{\ref{aff1}}
\and M.~Walmsley\orcid{0000-0002-6408-4181}\inst{\ref{aff2},\ref{aff3}}
\and P.~J.~Marshall\orcid{0000-0002-0113-5770}\inst{\ref{aff4},\ref{aff5}}
\and A.~More\inst{\ref{aff6}}
\and T.~E.~Collett\orcid{0000-0001-5564-3140}\inst{\ref{aff7}}
\and N.~E.~P.~Lines\orcid{0009-0004-7751-1914}\inst{\ref{aff7}}
\and L.~Leuzzi\orcid{0009-0006-4479-7017}\inst{\ref{aff8},\ref{aff9}}
\and A.~Manj\'on-Garc\'ia\orcid{0000-0002-7413-8825}\inst{\ref{aff10}}
\and S.~H.~Vincken\inst{\ref{aff11}}
\and J.~Wilde\orcid{0000-0002-4460-7379}\inst{\ref{aff12}}
\and R.~Pearce-Casey\inst{\ref{aff13}}
\and I.~T.~Andika\orcid{0000-0001-6102-9526}\inst{\ref{aff14},\ref{aff15}}
\and J.~A.~Acevedo~Barroso\orcid{0000-0002-9654-1711}\inst{\ref{aff16}}
\and T.~Li\orcid{0009-0005-5008-0381}\inst{\ref{aff7}}
\and A.~Melo\orcid{0000-0002-6449-3970}\inst{\ref{aff15},\ref{aff14}}
\and R.~B.~Metcalf\orcid{0000-0003-3167-2574}\inst{\ref{aff8},\ref{aff9}}
\and K.~Rojas\orcid{0000-0003-1391-6854}\inst{\ref{aff11},\ref{aff7}}
\and B.~Cl\'ement\orcid{0000-0002-7966-3661}\inst{\ref{aff16},\ref{aff17}}
\and H.~Degaudenzi\orcid{0000-0002-5887-6799}\inst{\ref{aff18}}
\and F.~Courbin\orcid{0000-0003-0758-6510}\inst{\ref{aff12},\ref{aff19}}
\and G.~Despali\orcid{0000-0001-6150-4112}\inst{\ref{aff8},\ref{aff9},\ref{aff20}}
\and R.~Gavazzi\orcid{0000-0002-5540-6935}\inst{\ref{aff21},\ref{aff22}}
\and S.~Schuldt\orcid{0000-0003-2497-6334}\inst{\ref{aff23},\ref{aff24}}
\and B.~C.~Nagam\orcid{0000-0002-3724-7694}\inst{\ref{aff25},\ref{aff26}}
\and D.~Sluse\orcid{0000-0001-6116-2095}\inst{\ref{aff27}}
\and C.~Tortora\orcid{0000-0001-7958-6531}\inst{\ref{aff28}}
\and H.~Dom\'inguez~S\'anchez\orcid{0000-0002-9013-1316}\inst{\ref{aff29}}
\and K.~Finner\orcid{0000-0002-4462-0709}\inst{\ref{aff30}}
\and A.~Galan\orcid{0000-0003-2547-9815}\inst{\ref{aff14},\ref{aff15}}
\and C.~Giocoli\orcid{0000-0002-9590-7961}\inst{\ref{aff9},\ref{aff20}}
\and L.~Guzzo\orcid{0000-0001-8264-5192}\inst{\ref{aff23},\ref{aff31},\ref{aff32}}
\and N.~B.~Hogg\orcid{0000-0001-9346-4477}\inst{\ref{aff33}}
\and K.~Jahnke\orcid{0000-0003-3804-2137}\inst{\ref{aff34}}
\and S.~Kruk\orcid{0000-0001-8010-8879}\inst{\ref{aff35}}
\and G.~Mahler\orcid{0000-0003-3266-2001}\inst{\ref{aff27},\ref{aff36},\ref{aff37}}
\and M.~Millon\orcid{0000-0001-7051-497X}\inst{\ref{aff38}}
\and P.~Nugent\orcid{0000-0002-3389-0586}\inst{\ref{aff39}}
\and J.~Pearson\orcid{0000-0001-8555-8561}\inst{\ref{aff13}}
\and L.~R.~Ecker\orcid{0009-0005-3508-2469}\inst{\ref{aff40},\ref{aff41}}
\and A.~Sainz~de~Murieta\inst{\ref{aff7}}
\and C.~Scarlata\orcid{0000-0002-9136-8876}\inst{\ref{aff25}}
\and S.~Serjeant\orcid{0000-0002-0517-7943}\inst{\ref{aff13}}
\and A.~Sonnenfeld\orcid{0000-0002-6061-5977}\inst{\ref{aff42}}
\and C.~Spiniello\orcid{0000-0002-3909-6359}\inst{\ref{aff1}}
\and T.~T.~Thai\orcid{0000-0002-8408-4816}\inst{\ref{aff43}}
\and L.~Ulivi\orcid{0009-0001-3291-5382}\inst{\ref{aff44},\ref{aff45},\ref{aff46}}
\and L.~Weisenbach\orcid{0000-0003-1175-8004}\inst{\ref{aff7}}
\and M.~Zumalacarregui\orcid{0000-0002-9943-6490}\inst{\ref{aff47}}
\and N.~Aghanim\orcid{0000-0002-6688-8992}\inst{\ref{aff48}}
\and B.~Altieri\orcid{0000-0003-3936-0284}\inst{\ref{aff35}}
\and A.~Amara\inst{\ref{aff49}}
\and S.~Andreon\orcid{0000-0002-2041-8784}\inst{\ref{aff31}}
\and N.~Auricchio\orcid{0000-0003-4444-8651}\inst{\ref{aff9}}
\and H.~Aussel\orcid{0000-0002-1371-5705}\inst{\ref{aff50}}
\and C.~Baccigalupi\orcid{0000-0002-8211-1630}\inst{\ref{aff51},\ref{aff52},\ref{aff53},\ref{aff54}}
\and M.~Baldi\orcid{0000-0003-4145-1943}\inst{\ref{aff55},\ref{aff9},\ref{aff20}}
\and A.~Balestra\orcid{0000-0002-6967-261X}\inst{\ref{aff56}}
\and S.~Bardelli\orcid{0000-0002-8900-0298}\inst{\ref{aff9}}
\and P.~Battaglia\orcid{0000-0002-7337-5909}\inst{\ref{aff9}}
\and R.~Bender\orcid{0000-0001-7179-0626}\inst{\ref{aff41},\ref{aff40}}
\and A.~Biviano\orcid{0000-0002-0857-0732}\inst{\ref{aff52},\ref{aff51}}
\and A.~Bonchi\orcid{0000-0002-2667-5482}\inst{\ref{aff57}}
\and E.~Branchini\orcid{0000-0002-0808-6908}\inst{\ref{aff58},\ref{aff59},\ref{aff31}}
\and M.~Brescia\orcid{0000-0001-9506-5680}\inst{\ref{aff60},\ref{aff28}}
\and J.~Brinchmann\orcid{0000-0003-4359-8797}\inst{\ref{aff61},\ref{aff62}}
\and S.~Camera\orcid{0000-0003-3399-3574}\inst{\ref{aff63},\ref{aff64},\ref{aff65}}
\and G.~Ca\~nas-Herrera\orcid{0000-0003-2796-2149}\inst{\ref{aff66},\ref{aff67},\ref{aff68}}
\and V.~Capobianco\orcid{0000-0002-3309-7692}\inst{\ref{aff65}}
\and C.~Carbone\orcid{0000-0003-0125-3563}\inst{\ref{aff24}}
\and V.~F.~Cardone\inst{\ref{aff69},\ref{aff70}}
\and J.~Carretero\orcid{0000-0002-3130-0204}\inst{\ref{aff71},\ref{aff72}}
\and M.~Castellano\orcid{0000-0001-9875-8263}\inst{\ref{aff69}}
\and G.~Castignani\orcid{0000-0001-6831-0687}\inst{\ref{aff9}}
\and S.~Cavuoti\orcid{0000-0002-3787-4196}\inst{\ref{aff28},\ref{aff73}}
\and K.~C.~Chambers\orcid{0000-0001-6965-7789}\inst{\ref{aff74}}
\and A.~Cimatti\inst{\ref{aff75}}
\and C.~Colodro-Conde\inst{\ref{aff76}}
\and G.~Congedo\orcid{0000-0003-2508-0046}\inst{\ref{aff77}}
\and C.~J.~Conselice\orcid{0000-0003-1949-7638}\inst{\ref{aff3}}
\and L.~Conversi\orcid{0000-0002-6710-8476}\inst{\ref{aff78},\ref{aff35}}
\and Y.~Copin\orcid{0000-0002-5317-7518}\inst{\ref{aff79}}
\and H.~M.~Courtois\orcid{0000-0003-0509-1776}\inst{\ref{aff80}}
\and M.~Cropper\orcid{0000-0003-4571-9468}\inst{\ref{aff81}}
\and A.~Da~Silva\orcid{0000-0002-6385-1609}\inst{\ref{aff82},\ref{aff83}}
\and G.~De~Lucia\orcid{0000-0002-6220-9104}\inst{\ref{aff52}}
\and A.~M.~Di~Giorgio\orcid{0000-0002-4767-2360}\inst{\ref{aff84}}
\and C.~Dolding\orcid{0009-0003-7199-6108}\inst{\ref{aff81}}
\and H.~Dole\orcid{0000-0002-9767-3839}\inst{\ref{aff48}}
\and F.~Dubath\orcid{0000-0002-6533-2810}\inst{\ref{aff18}}
\and C.~A.~J.~Duncan\orcid{0009-0003-3573-0791}\inst{\ref{aff3}}
\and X.~Dupac\inst{\ref{aff35}}
\and S.~Dusini\orcid{0000-0002-1128-0664}\inst{\ref{aff85}}
\and A.~Ealet\orcid{0000-0003-3070-014X}\inst{\ref{aff79}}
\and S.~Escoffier\orcid{0000-0002-2847-7498}\inst{\ref{aff86}}
\and M.~Farina\orcid{0000-0002-3089-7846}\inst{\ref{aff84}}
\and R.~Farinelli\inst{\ref{aff9}}
\and F.~Faustini\orcid{0000-0001-6274-5145}\inst{\ref{aff69},\ref{aff57}}
\and S.~Ferriol\inst{\ref{aff79}}
\and F.~Finelli\orcid{0000-0002-6694-3269}\inst{\ref{aff9},\ref{aff87}}
\and S.~Fotopoulou\orcid{0000-0002-9686-254X}\inst{\ref{aff88}}
\and M.~Frailis\orcid{0000-0002-7400-2135}\inst{\ref{aff52}}
\and E.~Franceschi\orcid{0000-0002-0585-6591}\inst{\ref{aff9}}
\and M.~Fumana\orcid{0000-0001-6787-5950}\inst{\ref{aff24}}
\and S.~Galeotta\orcid{0000-0002-3748-5115}\inst{\ref{aff52}}
\and K.~George\orcid{0000-0002-1734-8455}\inst{\ref{aff40}}
\and B.~Gillis\orcid{0000-0002-4478-1270}\inst{\ref{aff77}}
\and P.~G\'omez-Alvarez\orcid{0000-0002-8594-5358}\inst{\ref{aff89},\ref{aff35}}
\and J.~Gracia-Carpio\inst{\ref{aff41}}
\and B.~R.~Granett\orcid{0000-0003-2694-9284}\inst{\ref{aff31}}
\and A.~Grazian\orcid{0000-0002-5688-0663}\inst{\ref{aff56}}
\and F.~Grupp\inst{\ref{aff41},\ref{aff40}}
\and S.~V.~H.~Haugan\orcid{0000-0001-9648-7260}\inst{\ref{aff90}}
\and J.~Hoar\inst{\ref{aff35}}
\and W.~Holmes\inst{\ref{aff91}}
\and F.~Hormuth\inst{\ref{aff92}}
\and A.~Hornstrup\orcid{0000-0002-3363-0936}\inst{\ref{aff93},\ref{aff94}}
\and P.~Hudelot\inst{\ref{aff22}}
\and M.~Jhabvala\inst{\ref{aff95}}
\and B.~Joachimi\orcid{0000-0001-7494-1303}\inst{\ref{aff96}}
\and E.~Keih\"anen\orcid{0000-0003-1804-7715}\inst{\ref{aff97}}
\and S.~Kermiche\orcid{0000-0002-0302-5735}\inst{\ref{aff86}}
\and A.~Kiessling\orcid{0000-0002-2590-1273}\inst{\ref{aff91}}
\and B.~Kubik\orcid{0009-0006-5823-4880}\inst{\ref{aff79}}
\and M.~K\"ummel\orcid{0000-0003-2791-2117}\inst{\ref{aff40}}
\and M.~Kunz\orcid{0000-0002-3052-7394}\inst{\ref{aff98}}
\and H.~Kurki-Suonio\orcid{0000-0002-4618-3063}\inst{\ref{aff99},\ref{aff100}}
\and Q.~Le~Boulc'h\inst{\ref{aff101}}
\and A.~M.~C.~Le~Brun\orcid{0000-0002-0936-4594}\inst{\ref{aff102}}
\and D.~Le~Mignant\orcid{0000-0002-5339-5515}\inst{\ref{aff21}}
\and S.~Ligori\orcid{0000-0003-4172-4606}\inst{\ref{aff65}}
\and P.~B.~Lilje\orcid{0000-0003-4324-7794}\inst{\ref{aff90}}
\and V.~Lindholm\orcid{0000-0003-2317-5471}\inst{\ref{aff99},\ref{aff100}}
\and I.~Lloro\orcid{0000-0001-5966-1434}\inst{\ref{aff103}}
\and G.~Mainetti\orcid{0000-0003-2384-2377}\inst{\ref{aff101}}
\and D.~Maino\inst{\ref{aff23},\ref{aff24},\ref{aff32}}
\and E.~Maiorano\orcid{0000-0003-2593-4355}\inst{\ref{aff9}}
\and O.~Mansutti\orcid{0000-0001-5758-4658}\inst{\ref{aff52}}
\and S.~Marcin\inst{\ref{aff104}}
\and O.~Marggraf\orcid{0000-0001-7242-3852}\inst{\ref{aff105}}
\and M.~Martinelli\orcid{0000-0002-6943-7732}\inst{\ref{aff69},\ref{aff70}}
\and N.~Martinet\orcid{0000-0003-2786-7790}\inst{\ref{aff21}}
\and F.~Marulli\orcid{0000-0002-8850-0303}\inst{\ref{aff8},\ref{aff9},\ref{aff20}}
\and R.~Massey\orcid{0000-0002-6085-3780}\inst{\ref{aff37}}
\and S.~Maurogordato\inst{\ref{aff106}}
\and E.~Medinaceli\orcid{0000-0002-4040-7783}\inst{\ref{aff9}}
\and S.~Mei\orcid{0000-0002-2849-559X}\inst{\ref{aff107},\ref{aff108}}
\and M.~Melchior\inst{\ref{aff11}}
\and Y.~Mellier\inst{\ref{aff109},\ref{aff22}}
\and M.~Meneghetti\orcid{0000-0003-1225-7084}\inst{\ref{aff9},\ref{aff20}}
\and E.~Merlin\orcid{0000-0001-6870-8900}\inst{\ref{aff69}}
\and G.~Meylan\inst{\ref{aff16}}
\and A.~Mora\orcid{0000-0002-1922-8529}\inst{\ref{aff110}}
\and M.~Moresco\orcid{0000-0002-7616-7136}\inst{\ref{aff8},\ref{aff9}}
\and L.~Moscardini\orcid{0000-0002-3473-6716}\inst{\ref{aff8},\ref{aff9},\ref{aff20}}
\and R.~Nakajima\orcid{0009-0009-1213-7040}\inst{\ref{aff105}}
\and C.~Neissner\orcid{0000-0001-8524-4968}\inst{\ref{aff111},\ref{aff72}}
\and R.~C.~Nichol\orcid{0000-0003-0939-6518}\inst{\ref{aff49}}
\and S.-M.~Niemi\inst{\ref{aff66}}
\and J.~W.~Nightingale\orcid{0000-0002-8987-7401}\inst{\ref{aff112}}
\and C.~Padilla\orcid{0000-0001-7951-0166}\inst{\ref{aff111}}
\and S.~Paltani\orcid{0000-0002-8108-9179}\inst{\ref{aff18}}
\and F.~Pasian\orcid{0000-0002-4869-3227}\inst{\ref{aff52}}
\and K.~Pedersen\inst{\ref{aff113}}
\and W.~J.~Percival\orcid{0000-0002-0644-5727}\inst{\ref{aff114},\ref{aff115},\ref{aff116}}
\and V.~Pettorino\inst{\ref{aff66}}
\and S.~Pires\orcid{0000-0002-0249-2104}\inst{\ref{aff50}}
\and G.~Polenta\orcid{0000-0003-4067-9196}\inst{\ref{aff57}}
\and M.~Poncet\inst{\ref{aff117}}
\and L.~A.~Popa\inst{\ref{aff118}}
\and L.~Pozzetti\orcid{0000-0001-7085-0412}\inst{\ref{aff9}}
\and F.~Raison\orcid{0000-0002-7819-6918}\inst{\ref{aff41}}
\and R.~Rebolo\orcid{0000-0003-3767-7085}\inst{\ref{aff76},\ref{aff119},\ref{aff120}}
\and A.~Renzi\orcid{0000-0001-9856-1970}\inst{\ref{aff121},\ref{aff85}}
\and J.~Rhodes\orcid{0000-0002-4485-8549}\inst{\ref{aff91}}
\and G.~Riccio\inst{\ref{aff28}}
\and E.~Romelli\orcid{0000-0003-3069-9222}\inst{\ref{aff52}}
\and M.~Roncarelli\orcid{0000-0001-9587-7822}\inst{\ref{aff9}}
\and R.~Saglia\orcid{0000-0003-0378-7032}\inst{\ref{aff40},\ref{aff41}}
\and Z.~Sakr\orcid{0000-0002-4823-3757}\inst{\ref{aff122},\ref{aff123},\ref{aff124}}
\and D.~Sapone\orcid{0000-0001-7089-4503}\inst{\ref{aff125}}
\and B.~Sartoris\orcid{0000-0003-1337-5269}\inst{\ref{aff40},\ref{aff52}}
\and J.~A.~Schewtschenko\orcid{0000-0002-4913-6393}\inst{\ref{aff77}}
\and P.~Schneider\orcid{0000-0001-8561-2679}\inst{\ref{aff105}}
\and A.~Secroun\orcid{0000-0003-0505-3710}\inst{\ref{aff86}}
\and G.~Seidel\orcid{0000-0003-2907-353X}\inst{\ref{aff34}}
\and S.~Serrano\orcid{0000-0002-0211-2861}\inst{\ref{aff126},\ref{aff127},\ref{aff128}}
\and P.~Simon\inst{\ref{aff105}}
\and C.~Sirignano\orcid{0000-0002-0995-7146}\inst{\ref{aff121},\ref{aff85}}
\and G.~Sirri\orcid{0000-0003-2626-2853}\inst{\ref{aff20}}
\and L.~Stanco\orcid{0000-0002-9706-5104}\inst{\ref{aff85}}
\and J.~Steinwagner\orcid{0000-0001-7443-1047}\inst{\ref{aff41}}
\and P.~Tallada-Cresp\'{i}\orcid{0000-0002-1336-8328}\inst{\ref{aff71},\ref{aff72}}
\and A.~N.~Taylor\inst{\ref{aff77}}
\and I.~Tereno\inst{\ref{aff82},\ref{aff129}}
\and S.~Toft\orcid{0000-0003-3631-7176}\inst{\ref{aff130},\ref{aff131}}
\and R.~Toledo-Moreo\orcid{0000-0002-2997-4859}\inst{\ref{aff132}}
\and F.~Torradeflot\orcid{0000-0003-1160-1517}\inst{\ref{aff72},\ref{aff71}}
\and I.~Tutusaus\orcid{0000-0002-3199-0399}\inst{\ref{aff123}}
\and E.~A.~Valentijn\inst{\ref{aff26}}
\and L.~Valenziano\orcid{0000-0002-1170-0104}\inst{\ref{aff9},\ref{aff87}}
\and J.~Valiviita\orcid{0000-0001-6225-3693}\inst{\ref{aff99},\ref{aff100}}
\and T.~Vassallo\orcid{0000-0001-6512-6358}\inst{\ref{aff40},\ref{aff52}}
\and G.~Verdoes~Kleijn\orcid{0000-0001-5803-2580}\inst{\ref{aff26}}
\and A.~Veropalumbo\orcid{0000-0003-2387-1194}\inst{\ref{aff31},\ref{aff59},\ref{aff58}}
\and Y.~Wang\orcid{0000-0002-4749-2984}\inst{\ref{aff133}}
\and J.~Weller\orcid{0000-0002-8282-2010}\inst{\ref{aff40},\ref{aff41}}
\and A.~Zacchei\orcid{0000-0003-0396-1192}\inst{\ref{aff52},\ref{aff51}}
\and G.~Zamorani\orcid{0000-0002-2318-301X}\inst{\ref{aff9}}
\and F.~M.~Zerbi\inst{\ref{aff31}}
\and E.~Zucca\orcid{0000-0002-5845-8132}\inst{\ref{aff9}}
\and V.~Allevato\orcid{0000-0001-7232-5152}\inst{\ref{aff28}}
\and M.~Ballardini\orcid{0000-0003-4481-3559}\inst{\ref{aff134},\ref{aff135},\ref{aff9}}
\and M.~Bolzonella\orcid{0000-0003-3278-4607}\inst{\ref{aff9}}
\and E.~Bozzo\orcid{0000-0002-8201-1525}\inst{\ref{aff18}}
\and C.~Burigana\orcid{0000-0002-3005-5796}\inst{\ref{aff136},\ref{aff87}}
\and R.~Cabanac\orcid{0000-0001-6679-2600}\inst{\ref{aff123}}
\and A.~Cappi\inst{\ref{aff9},\ref{aff106}}
\and D.~Di~Ferdinando\inst{\ref{aff20}}
\and J.~A.~Escartin~Vigo\inst{\ref{aff41}}
\and L.~Gabarra\orcid{0000-0002-8486-8856}\inst{\ref{aff1}}
\and M.~Huertas-Company\orcid{0000-0002-1416-8483}\inst{\ref{aff76},\ref{aff137},\ref{aff138},\ref{aff139}}
\and J.~Mart\'{i}n-Fleitas\orcid{0000-0002-8594-569X}\inst{\ref{aff110}}
\and S.~Matthew\orcid{0000-0001-8448-1697}\inst{\ref{aff77}}
\and N.~Mauri\orcid{0000-0001-8196-1548}\inst{\ref{aff75},\ref{aff20}}
\and A.~A.~Nucita\inst{\ref{aff140},\ref{aff141},\ref{aff142}}
\and A.~Pezzotta\orcid{0000-0003-0726-2268}\inst{\ref{aff143},\ref{aff41}}
\and M.~P\"ontinen\orcid{0000-0001-5442-2530}\inst{\ref{aff99}}
\and C.~Porciani\orcid{0000-0002-7797-2508}\inst{\ref{aff105}}
\and I.~Risso\orcid{0000-0003-2525-7761}\inst{\ref{aff144}}
\and V.~Scottez\inst{\ref{aff109},\ref{aff145}}
\and M.~Sereno\orcid{0000-0003-0302-0325}\inst{\ref{aff9},\ref{aff20}}
\and M.~Tenti\orcid{0000-0002-4254-5901}\inst{\ref{aff20}}
\and M.~Viel\orcid{0000-0002-2642-5707}\inst{\ref{aff51},\ref{aff52},\ref{aff54},\ref{aff53},\ref{aff146}}
\and M.~Wiesmann\orcid{0009-0000-8199-5860}\inst{\ref{aff90}}
\and Y.~Akrami\orcid{0000-0002-2407-7956}\inst{\ref{aff147},\ref{aff148}}
\and S.~Anselmi\orcid{0000-0002-3579-9583}\inst{\ref{aff85},\ref{aff121},\ref{aff149}}
\and M.~Archidiacono\orcid{0000-0003-4952-9012}\inst{\ref{aff23},\ref{aff32}}
\and F.~Atrio-Barandela\orcid{0000-0002-2130-2513}\inst{\ref{aff150}}
\and C.~Benoist\inst{\ref{aff106}}
\and K.~Benson\inst{\ref{aff81}}
\and P.~Bergamini\orcid{0000-0003-1383-9414}\inst{\ref{aff23},\ref{aff9}}
\and D.~Bertacca\orcid{0000-0002-2490-7139}\inst{\ref{aff121},\ref{aff56},\ref{aff85}}
\and M.~Bethermin\orcid{0000-0002-3915-2015}\inst{\ref{aff151}}
\and A.~Blanchard\orcid{0000-0001-8555-9003}\inst{\ref{aff123}}
\and L.~Blot\orcid{0000-0002-9622-7167}\inst{\ref{aff152},\ref{aff149}}
\and S.~Borgani\orcid{0000-0001-6151-6439}\inst{\ref{aff153},\ref{aff51},\ref{aff52},\ref{aff53},\ref{aff146}}
\and M.~L.~Brown\orcid{0000-0002-0370-8077}\inst{\ref{aff3}}
\and S.~Bruton\orcid{0000-0002-6503-5218}\inst{\ref{aff154}}
\and A.~Calabro\orcid{0000-0003-2536-1614}\inst{\ref{aff69}}
\and F.~Caro\inst{\ref{aff69}}
\and C.~S.~Carvalho\inst{\ref{aff129}}
\and T.~Castro\orcid{0000-0002-6292-3228}\inst{\ref{aff52},\ref{aff53},\ref{aff51},\ref{aff146}}
\and F.~Cogato\orcid{0000-0003-4632-6113}\inst{\ref{aff8},\ref{aff9}}
\and A.~R.~Cooray\orcid{0000-0002-3892-0190}\inst{\ref{aff155}}
\and O.~Cucciati\orcid{0000-0002-9336-7551}\inst{\ref{aff9}}
\and S.~Davini\orcid{0000-0003-3269-1718}\inst{\ref{aff59}}
\and F.~De~Paolis\orcid{0000-0001-6460-7563}\inst{\ref{aff140},\ref{aff141},\ref{aff142}}
\and G.~Desprez\orcid{0000-0001-8325-1742}\inst{\ref{aff26}}
\and A.~D\'iaz-S\'anchez\orcid{0000-0003-0748-4768}\inst{\ref{aff10}}
\and J.~J.~Diaz\inst{\ref{aff137}}
\and S.~Di~Domizio\orcid{0000-0003-2863-5895}\inst{\ref{aff58},\ref{aff59}}
\and J.~M.~Diego\orcid{0000-0001-9065-3926}\inst{\ref{aff29}}
\and P.-A.~Duc\orcid{0000-0003-3343-6284}\inst{\ref{aff151}}
\and A.~Enia\orcid{0000-0002-0200-2857}\inst{\ref{aff55},\ref{aff9}}
\and Y.~Fang\inst{\ref{aff40}}
\and A.~G.~Ferrari\orcid{0009-0005-5266-4110}\inst{\ref{aff20}}
\and P.~G.~Ferreira\orcid{0000-0002-3021-2851}\inst{\ref{aff1}}
\and A.~Finoguenov\orcid{0000-0002-4606-5403}\inst{\ref{aff99}}
\and A.~Fontana\orcid{0000-0003-3820-2823}\inst{\ref{aff69}}
\and A.~Franco\orcid{0000-0002-4761-366X}\inst{\ref{aff141},\ref{aff140},\ref{aff142}}
\and K.~Ganga\orcid{0000-0001-8159-8208}\inst{\ref{aff107}}
\and J.~Garc\'ia-Bellido\orcid{0000-0002-9370-8360}\inst{\ref{aff147}}
\and T.~Gasparetto\orcid{0000-0002-7913-4866}\inst{\ref{aff52}}
\and V.~Gautard\inst{\ref{aff156}}
\and E.~Gaztanaga\orcid{0000-0001-9632-0815}\inst{\ref{aff128},\ref{aff126},\ref{aff7}}
\and F.~Giacomini\orcid{0000-0002-3129-2814}\inst{\ref{aff20}}
\and F.~Gianotti\orcid{0000-0003-4666-119X}\inst{\ref{aff9}}
\and G.~Gozaliasl\orcid{0000-0002-0236-919X}\inst{\ref{aff157},\ref{aff99}}
\and C.~M.~Gutierrez\orcid{0000-0001-7854-783X}\inst{\ref{aff158}}
\and A.~Hall\orcid{0000-0002-3139-8651}\inst{\ref{aff77}}
\and W.~G.~Hartley\inst{\ref{aff18}}
\and C.~Hern\'andez-Monteagudo\orcid{0000-0001-5471-9166}\inst{\ref{aff120},\ref{aff76}}
\and H.~Hildebrandt\orcid{0000-0002-9814-3338}\inst{\ref{aff159}}
\and J.~Hjorth\orcid{0000-0002-4571-2306}\inst{\ref{aff113}}
\and J.~J.~E.~Kajava\orcid{0000-0002-3010-8333}\inst{\ref{aff160},\ref{aff161}}
\and Y.~Kang\orcid{0009-0000-8588-7250}\inst{\ref{aff18}}
\and V.~Kansal\orcid{0000-0002-4008-6078}\inst{\ref{aff162},\ref{aff163}}
\and D.~Karagiannis\orcid{0000-0002-4927-0816}\inst{\ref{aff134},\ref{aff164}}
\and K.~Kiiveri\inst{\ref{aff97}}
\and C.~C.~Kirkpatrick\inst{\ref{aff97}}
\and J.~Le~Graet\orcid{0000-0001-6523-7971}\inst{\ref{aff86}}
\and L.~Legrand\orcid{0000-0003-0610-5252}\inst{\ref{aff165},\ref{aff166}}
\and M.~Lembo\orcid{0000-0002-5271-5070}\inst{\ref{aff134},\ref{aff135}}
\and F.~Lepori\orcid{0009-0000-5061-7138}\inst{\ref{aff167}}
\and G.~Leroy\orcid{0009-0004-2523-4425}\inst{\ref{aff36},\ref{aff37}}
\and G.~F.~Lesci\orcid{0000-0002-4607-2830}\inst{\ref{aff8},\ref{aff9}}
\and J.~Lesgourgues\orcid{0000-0001-7627-353X}\inst{\ref{aff168}}
\and T.~I.~Liaudat\orcid{0000-0002-9104-314X}\inst{\ref{aff169}}
\and A.~Loureiro\orcid{0000-0002-4371-0876}\inst{\ref{aff170},\ref{aff171}}
\and J.~Macias-Perez\orcid{0000-0002-5385-2763}\inst{\ref{aff172}}
\and G.~Maggio\orcid{0000-0003-4020-4836}\inst{\ref{aff52}}
\and M.~Magliocchetti\orcid{0000-0001-9158-4838}\inst{\ref{aff84}}
\and E.~A.~Magnier\orcid{0000-0002-7965-2815}\inst{\ref{aff74}}
\and F.~Mannucci\orcid{0000-0002-4803-2381}\inst{\ref{aff46}}
\and R.~Maoli\orcid{0000-0002-6065-3025}\inst{\ref{aff173},\ref{aff69}}
\and C.~J.~A.~P.~Martins\orcid{0000-0002-4886-9261}\inst{\ref{aff174},\ref{aff61}}
\and L.~Maurin\orcid{0000-0002-8406-0857}\inst{\ref{aff48}}
\and M.~Miluzio\inst{\ref{aff35},\ref{aff175}}
\and P.~Monaco\orcid{0000-0003-2083-7564}\inst{\ref{aff153},\ref{aff52},\ref{aff53},\ref{aff51}}
\and A.~Montoro\orcid{0000-0003-4730-8590}\inst{\ref{aff128},\ref{aff126}}
\and C.~Moretti\orcid{0000-0003-3314-8936}\inst{\ref{aff54},\ref{aff146},\ref{aff52},\ref{aff51},\ref{aff53}}
\and G.~Morgante\inst{\ref{aff9}}
\and K.~Naidoo\orcid{0000-0002-9182-1802}\inst{\ref{aff7}}
\and A.~Navarro-Alsina\orcid{0000-0002-3173-2592}\inst{\ref{aff105}}
\and S.~Nesseris\orcid{0000-0002-0567-0324}\inst{\ref{aff147}}
\and F.~Passalacqua\orcid{0000-0002-8606-4093}\inst{\ref{aff121},\ref{aff85}}
\and K.~Paterson\orcid{0000-0001-8340-3486}\inst{\ref{aff34}}
\and L.~Patrizii\inst{\ref{aff20}}
\and A.~Pisani\orcid{0000-0002-6146-4437}\inst{\ref{aff86},\ref{aff176}}
\and D.~Potter\orcid{0000-0002-0757-5195}\inst{\ref{aff167}}
\and S.~Quai\orcid{0000-0002-0449-8163}\inst{\ref{aff8},\ref{aff9}}
\and M.~Radovich\orcid{0000-0002-3585-866X}\inst{\ref{aff56}}
\and S.~Sacquegna\orcid{0000-0002-8433-6630}\inst{\ref{aff140},\ref{aff141},\ref{aff142}}
\and M.~Sahl\'en\orcid{0000-0003-0973-4804}\inst{\ref{aff177}}
\and D.~B.~Sanders\orcid{0000-0002-1233-9998}\inst{\ref{aff74}}
\and E.~Sarpa\orcid{0000-0002-1256-655X}\inst{\ref{aff54},\ref{aff146},\ref{aff53}}
\and A.~Schneider\orcid{0000-0001-7055-8104}\inst{\ref{aff167}}
\and D.~Sciotti\orcid{0009-0008-4519-2620}\inst{\ref{aff69},\ref{aff70}}
\and E.~Sellentin\inst{\ref{aff178},\ref{aff68}}
\and L.~C.~Smith\orcid{0000-0002-3259-2771}\inst{\ref{aff179}}
\and K.~Tanidis\orcid{0000-0001-9843-5130}\inst{\ref{aff1}}
\and G.~Testera\inst{\ref{aff59}}
\and R.~Teyssier\orcid{0000-0001-7689-0933}\inst{\ref{aff176}}
\and S.~Tosi\orcid{0000-0002-7275-9193}\inst{\ref{aff58},\ref{aff59},\ref{aff31}}
\and A.~Troja\orcid{0000-0003-0239-4595}\inst{\ref{aff121},\ref{aff85}}
\and M.~Tucci\inst{\ref{aff18}}
\and C.~Valieri\inst{\ref{aff20}}
\and A.~Venhola\orcid{0000-0001-6071-4564}\inst{\ref{aff180}}
\and D.~Vergani\orcid{0000-0003-0898-2216}\inst{\ref{aff9}}
\and G.~Vernardos\orcid{0000-0001-8554-7248}\inst{\ref{aff181},\ref{aff182}}
\and G.~Verza\orcid{0000-0002-1886-8348}\inst{\ref{aff183}}
\and P.~Vielzeuf\orcid{0000-0003-2035-9339}\inst{\ref{aff86}}
\and N.~A.~Walton\orcid{0000-0003-3983-8778}\inst{\ref{aff179}}
\and D.~Scott\orcid{0000-0002-6878-9840}\inst{\ref{aff184}}}
										   
\institute{Department of Physics, Oxford University, Keble Road, Oxford OX1 3RH, UK\label{aff1}
\and
David A. Dunlap Department of Astronomy \& Astrophysics, University of Toronto, 50 St George Street, Toronto, Ontario M5S 3H4, Canada\label{aff2}
\and
Jodrell Bank Centre for Astrophysics, Department of Physics and Astronomy, University of Manchester, Oxford Road, Manchester M13 9PL, UK\label{aff3}
\and
Kavli Institute for Particle Astrophysics \& Cosmology (KIPAC), Stanford University, Stanford, CA 94305, USA\label{aff4}
\and
SLAC National Accelerator Laboratory, 2575 Sand Hill Road, Menlo Park, CA 94025, USA\label{aff5}
\and
The Inter-University Centre for Astronomy and Astrophysics, Post Bag 4, Ganeshkhind, Pune 411007, India\label{aff6}
\and
Institute of Cosmology and Gravitation, University of Portsmouth, Portsmouth PO1 3FX, UK\label{aff7}
\and
Dipartimento di Fisica e Astronomia "Augusto Righi" - Alma Mater Studiorum Universit\`a di Bologna, via Piero Gobetti 93/2, 40129 Bologna, Italy\label{aff8}
\and
INAF-Osservatorio di Astrofisica e Scienza dello Spazio di Bologna, Via Piero Gobetti 93/3, 40129 Bologna, Italy\label{aff9}
\and
Departamento F\'isica Aplicada, Universidad Polit\'ecnica de Cartagena, Campus Muralla del Mar, 30202 Cartagena, Murcia, Spain\label{aff10}
\and
University of Applied Sciences and Arts of Northwestern Switzerland, School of Engineering, 5210 Windisch, Switzerland\label{aff11}
\and
Institut de Ci\`{e}ncies del Cosmos (ICCUB), Universitat de Barcelona (IEEC-UB), Mart\'{i} i Franqu\`{e}s 1, 08028 Barcelona, Spain\label{aff12}
\and
School of Physical Sciences, The Open University, Milton Keynes, MK7 6AA, UK\label{aff13}
\and
Technical University of Munich, TUM School of Natural Sciences, Physics Department, James-Franck-Str.~1, 85748 Garching, Germany\label{aff14}
\and
Max-Planck-Institut f\"ur Astrophysik, Karl-Schwarzschild-Str.~1, 85748 Garching, Germany\label{aff15}
\and
Institute of Physics, Laboratory of Astrophysics, Ecole Polytechnique F\'ed\'erale de Lausanne (EPFL), Observatoire de Sauverny, 1290 Versoix, Switzerland\label{aff16}
\and
SCITAS, Ecole Polytechnique F\'ed\'erale de Lausanne (EPFL), 1015 Lausanne, Switzerland\label{aff17}
\and
Department of Astronomy, University of Geneva, ch. d'Ecogia 16, 1290 Versoix, Switzerland\label{aff18}
\and
Instituci\'o Catalana de Recerca i Estudis Avan\c{c}ats (ICREA), Passeig de Llu\'{\i}s Companys 23, 08010 Barcelona, Spain\label{aff19}
\and
INFN-Sezione di Bologna, Viale Berti Pichat 6/2, 40127 Bologna, Italy\label{aff20}
\and
Aix-Marseille Universit\'e, CNRS, CNES, LAM, Marseille, France\label{aff21}
\and
Institut d'Astrophysique de Paris, UMR 7095, CNRS, and Sorbonne Universit\'e, 98 bis boulevard Arago, 75014 Paris, France\label{aff22}
\and
Dipartimento di Fisica "Aldo Pontremoli", Universit\`a degli Studi di Milano, Via Celoria 16, 20133 Milano, Italy\label{aff23}
\and
INAF-IASF Milano, Via Alfonso Corti 12, 20133 Milano, Italy\label{aff24}
\and
Minnesota Institute for Astrophysics, University of Minnesota, 116 Church St SE, Minneapolis, MN 55455, USA\label{aff25}
\and
Kapteyn Astronomical Institute, University of Groningen, PO Box 800, 9700 AV Groningen, The Netherlands\label{aff26}
\and
STAR Institute, University of Li{\`e}ge, Quartier Agora, All\'ee du six Ao\^ut 19c, 4000 Li\`ege, Belgium\label{aff27}
\and
INAF-Osservatorio Astronomico di Capodimonte, Via Moiariello 16, 80131 Napoli, Italy\label{aff28}
\and
Instituto de F\'isica de Cantabria, Edificio Juan Jord\'a, Avenida de los Castros, 39005 Santander, Spain\label{aff29}
\and
Caltech/IPAC, 1200 E. California Blvd., Pasadena, CA 91125, USA\label{aff30}
\and
INAF-Osservatorio Astronomico di Brera, Via Brera 28, 20122 Milano, Italy\label{aff31}
\and
INFN-Sezione di Milano, Via Celoria 16, 20133 Milano, Italy\label{aff32}
\and
Laboratoire univers et particules de Montpellier, Universit\'e de Montpellier, CNRS, 34090 Montpellier, France\label{aff33}
\and
Max-Planck-Institut f\"ur Astronomie, K\"onigstuhl 17, 69117 Heidelberg, Germany\label{aff34}
\and
ESAC/ESA, Camino Bajo del Castillo, s/n., Urb. Villafranca del Castillo, 28692 Villanueva de la Ca\~nada, Madrid, Spain\label{aff35}
\and
Department of Physics, Centre for Extragalactic Astronomy, Durham University, South Road, Durham, DH1 3LE, UK\label{aff36}
\and
Department of Physics, Institute for Computational Cosmology, Durham University, South Road, Durham, DH1 3LE, UK\label{aff37}
\and
Institute for Particle Physics and Astrophysics, Dept. of Physics, ETH Zurich, Wolfgang-Pauli-Strasse 27, 8093 Zurich, Switzerland\label{aff38}
\and
Lawrence Berkeley National Laboratory, One Cyclotron Road, Berkeley, CA 94720, USA\label{aff39}
\and
Universit\"ats-Sternwarte M\"unchen, Fakult\"at f\"ur Physik, Ludwig-Maximilians-Universit\"at M\"unchen, Scheinerstrasse 1, 81679 M\"unchen, Germany\label{aff40}
\and
Max Planck Institute for Extraterrestrial Physics, Giessenbachstr. 1, 85748 Garching, Germany\label{aff41}
\and
Department of Astronomy, School of Physics and Astronomy, Shanghai Jiao Tong University, Shanghai 200240, China\label{aff42}
\and
National Astronomical Observatory of Japan, 2-21-1 Osawa, Mitaka, Tokyo 181-8588, Japan\label{aff43}
\and
University of Trento, Via Sommarive 14, I-38123 Trento, Italy\label{aff44}
\and
Dipartimento di Fisica e Astronomia, Universit\`{a} di Firenze, via G. Sansone 1, 50019 Sesto Fiorentino, Firenze, Italy\label{aff45}
\and
INAF-Osservatorio Astrofisico di Arcetri, Largo E. Fermi 5, 50125, Firenze, Italy\label{aff46}
\and
Max Planck Institute for Gravitational Physics (Albert Einstein Institute), Am Muhlenberg 1, D-14476 Potsdam-Golm, Germany\label{aff47}
\and
Universit\'e Paris-Saclay, CNRS, Institut d'astrophysique spatiale, 91405, Orsay, France\label{aff48}
\and
School of Mathematics and Physics, University of Surrey, Guildford, Surrey, GU2 7XH, UK\label{aff49}
\and
Universit\'e Paris-Saclay, Universit\'e Paris Cit\'e, CEA, CNRS, AIM, 91191, Gif-sur-Yvette, France\label{aff50}
\and
IFPU, Institute for Fundamental Physics of the Universe, via Beirut 2, 34151 Trieste, Italy\label{aff51}
\and
INAF-Osservatorio Astronomico di Trieste, Via G. B. Tiepolo 11, 34143 Trieste, Italy\label{aff52}
\and
INFN, Sezione di Trieste, Via Valerio 2, 34127 Trieste TS, Italy\label{aff53}
\and
SISSA, International School for Advanced Studies, Via Bonomea 265, 34136 Trieste TS, Italy\label{aff54}
\and
Dipartimento di Fisica e Astronomia, Universit\`a di Bologna, Via Gobetti 93/2, 40129 Bologna, Italy\label{aff55}
\and
INAF-Osservatorio Astronomico di Padova, Via dell'Osservatorio 5, 35122 Padova, Italy\label{aff56}
\and
Space Science Data Center, Italian Space Agency, via del Politecnico snc, 00133 Roma, Italy\label{aff57}
\and
Dipartimento di Fisica, Universit\`a di Genova, Via Dodecaneso 33, 16146, Genova, Italy\label{aff58}
\and
INFN-Sezione di Genova, Via Dodecaneso 33, 16146, Genova, Italy\label{aff59}
\and
Department of Physics "E. Pancini", University Federico II, Via Cinthia 6, 80126, Napoli, Italy\label{aff60}
\and
Instituto de Astrof\'isica e Ci\^encias do Espa\c{c}o, Universidade do Porto, CAUP, Rua das Estrelas, PT4150-762 Porto, Portugal\label{aff61}
\and
Faculdade de Ci\^encias da Universidade do Porto, Rua do Campo de Alegre, 4150-007 Porto, Portugal\label{aff62}
\and
Dipartimento di Fisica, Universit\`a degli Studi di Torino, Via P. Giuria 1, 10125 Torino, Italy\label{aff63}
\and
INFN-Sezione di Torino, Via P. Giuria 1, 10125 Torino, Italy\label{aff64}
\and
INAF-Osservatorio Astrofisico di Torino, Via Osservatorio 20, 10025 Pino Torinese (TO), Italy\label{aff65}
\and
European Space Agency/ESTEC, Keplerlaan 1, 2201 AZ Noordwijk, The Netherlands\label{aff66}
\and
Institute Lorentz, Leiden University, Niels Bohrweg 2, 2333 CA Leiden, The Netherlands\label{aff67}
\and
Leiden Observatory, Leiden University, Einsteinweg 55, 2333 CC Leiden, The Netherlands\label{aff68}
\and
INAF-Osservatorio Astronomico di Roma, Via Frascati 33, 00078 Monteporzio Catone, Italy\label{aff69}
\and
INFN-Sezione di Roma, Piazzale Aldo Moro, 2 - c/o Dipartimento di Fisica, Edificio G. Marconi, 00185 Roma, Italy\label{aff70}
\and
Centro de Investigaciones Energ\'eticas, Medioambientales y Tecnol\'ogicas (CIEMAT), Avenida Complutense 40, 28040 Madrid, Spain\label{aff71}
\and
Port d'Informaci\'{o} Cient\'{i}fica, Campus UAB, C. Albareda s/n, 08193 Bellaterra (Barcelona), Spain\label{aff72}
\and
INFN section of Naples, Via Cinthia 6, 80126, Napoli, Italy\label{aff73}
\and
Institute for Astronomy, University of Hawaii, 2680 Woodlawn Drive, Honolulu, HI 96822, USA\label{aff74}
\and
Dipartimento di Fisica e Astronomia "Augusto Righi" - Alma Mater Studiorum Universit\`a di Bologna, Viale Berti Pichat 6/2, 40127 Bologna, Italy\label{aff75}
\and
Instituto de Astrof\'{\i}sica de Canarias, V\'{\i}a L\'actea, 38205 La Laguna, Tenerife, Spain\label{aff76}
\and
Institute for Astronomy, University of Edinburgh, Royal Observatory, Blackford Hill, Edinburgh EH9 3HJ, UK\label{aff77}
\and
European Space Agency/ESRIN, Largo Galileo Galilei 1, 00044 Frascati, Roma, Italy\label{aff78}
\and
Universit\'e Claude Bernard Lyon 1, CNRS/IN2P3, IP2I Lyon, UMR 5822, Villeurbanne, F-69100, France\label{aff79}
\and
UCB Lyon 1, CNRS/IN2P3, IUF, IP2I Lyon, 4 rue Enrico Fermi, 69622 Villeurbanne, France\label{aff80}
\and
Mullard Space Science Laboratory, University College London, Holmbury St Mary, Dorking, Surrey RH5 6NT, UK\label{aff81}
\and
Departamento de F\'isica, Faculdade de Ci\^encias, Universidade de Lisboa, Edif\'icio C8, Campo Grande, PT1749-016 Lisboa, Portugal\label{aff82}
\and
Instituto de Astrof\'isica e Ci\^encias do Espa\c{c}o, Faculdade de Ci\^encias, Universidade de Lisboa, Campo Grande, 1749-016 Lisboa, Portugal\label{aff83}
\and
INAF-Istituto di Astrofisica e Planetologia Spaziali, via del Fosso del Cavaliere, 100, 00100 Roma, Italy\label{aff84}
\and
INFN-Padova, Via Marzolo 8, 35131 Padova, Italy\label{aff85}
\and
Aix-Marseille Universit\'e, CNRS/IN2P3, CPPM, Marseille, France\label{aff86}
\and
INFN-Bologna, Via Irnerio 46, 40126 Bologna, Italy\label{aff87}
\and
School of Physics, HH Wills Physics Laboratory, University of Bristol, Tyndall Avenue, Bristol, BS8 1TL, UK\label{aff88}
\and
FRACTAL S.L.N.E., calle Tulip\'an 2, Portal 13 1A, 28231, Las Rozas de Madrid, Spain\label{aff89}
\and
Institute of Theoretical Astrophysics, University of Oslo, P.O. Box 1029 Blindern, 0315 Oslo, Norway\label{aff90}
\and
Jet Propulsion Laboratory, California Institute of Technology, 4800 Oak Grove Drive, Pasadena, CA, 91109, USA\label{aff91}
\and
Felix Hormuth Engineering, Goethestr. 17, 69181 Leimen, Germany\label{aff92}
\and
Technical University of Denmark, Elektrovej 327, 2800 Kgs. Lyngby, Denmark\label{aff93}
\and
Cosmic Dawn Center (DAWN), Denmark\label{aff94}
\and
NASA Goddard Space Flight Center, Greenbelt, MD 20771, USA\label{aff95}
\and
Department of Physics and Astronomy, University College London, Gower Street, London WC1E 6BT, UK\label{aff96}
\and
Department of Physics and Helsinki Institute of Physics, Gustaf H\"allstr\"omin katu 2, 00014 University of Helsinki, Finland\label{aff97}
\and
Universit\'e de Gen\`eve, D\'epartement de Physique Th\'eorique and Centre for Astroparticle Physics, 24 quai Ernest-Ansermet, CH-1211 Gen\`eve 4, Switzerland\label{aff98}
\and
Department of Physics, P.O. Box 64, 00014 University of Helsinki, Finland\label{aff99}
\and
Helsinki Institute of Physics, Gustaf H{\"a}llstr{\"o}min katu 2, University of Helsinki, Helsinki, Finland\label{aff100}
\and
Centre de Calcul de l'IN2P3/CNRS, 21 avenue Pierre de Coubertin 69627 Villeurbanne Cedex, France\label{aff101}
\and
Laboratoire d'etude de l'Univers et des phenomenes eXtremes, Observatoire de Paris, Universit\'e PSL, Sorbonne Universit\'e, CNRS, 92190 Meudon, France\label{aff102}
\and
SKA Observatory, Jodrell Bank, Lower Withington, Macclesfield, Cheshire SK11 9FT, UK\label{aff103}
\and
University of Applied Sciences and Arts of Northwestern Switzerland, School of Computer Science, 5210 Windisch, Switzerland\label{aff104}
\and
Universit\"at Bonn, Argelander-Institut f\"ur Astronomie, Auf dem H\"ugel 71, 53121 Bonn, Germany\label{aff105}
\and
Universit\'e C\^{o}te d'Azur, Observatoire de la C\^{o}te d'Azur, CNRS, Laboratoire Lagrange, Bd de l'Observatoire, CS 34229, 06304 Nice cedex 4, France\label{aff106}
\and
Universit\'e Paris Cit\'e, CNRS, Astroparticule et Cosmologie, 75013 Paris, France\label{aff107}
\and
CNRS-UCB International Research Laboratory, Centre Pierre Binetruy, IRL2007, CPB-IN2P3, Berkeley, USA\label{aff108}
\and
Institut d'Astrophysique de Paris, 98bis Boulevard Arago, 75014, Paris, France\label{aff109}
\and
Aurora Technology for European Space Agency (ESA), Camino bajo del Castillo, s/n, Urbanizacion Villafranca del Castillo, Villanueva de la Ca\~nada, 28692 Madrid, Spain\label{aff110}
\and
Institut de F\'{i}sica d'Altes Energies (IFAE), The Barcelona Institute of Science and Technology, Campus UAB, 08193 Bellaterra (Barcelona), Spain\label{aff111}
\and
School of Mathematics, Statistics and Physics, Newcastle University, Herschel Building, Newcastle-upon-Tyne, NE1 7RU, UK\label{aff112}
\and
DARK, Niels Bohr Institute, University of Copenhagen, Jagtvej 155, 2200 Copenhagen, Denmark\label{aff113}
\and
Waterloo Centre for Astrophysics, University of Waterloo, Waterloo, Ontario N2L 3G1, Canada\label{aff114}
\and
Department of Physics and Astronomy, University of Waterloo, Waterloo, Ontario N2L 3G1, Canada\label{aff115}
\and
Perimeter Institute for Theoretical Physics, Waterloo, Ontario N2L 2Y5, Canada\label{aff116}
\and
Centre National d'Etudes Spatiales -- Centre spatial de Toulouse, 18 avenue Edouard Belin, 31401 Toulouse Cedex 9, France\label{aff117}
\and
Institute of Space Science, Str. Atomistilor, nr. 409 M\u{a}gurele, Ilfov, 077125, Romania\label{aff118}
\and
Consejo Superior de Investigaciones Cientificas, Calle Serrano 117, 28006 Madrid, Spain\label{aff119}
\and
Universidad de La Laguna, Departamento de Astrof\'{\i}sica, 38206 La Laguna, Tenerife, Spain\label{aff120}
\and
Dipartimento di Fisica e Astronomia "G. Galilei", Universit\`a di Padova, Via Marzolo 8, 35131 Padova, Italy\label{aff121}
\and
Institut f\"ur Theoretische Physik, University of Heidelberg, Philosophenweg 16, 69120 Heidelberg, Germany\label{aff122}
\and
Institut de Recherche en Astrophysique et Plan\'etologie (IRAP), Universit\'e de Toulouse, CNRS, UPS, CNES, 14 Av. Edouard Belin, 31400 Toulouse, France\label{aff123}
\and
Universit\'e St Joseph; Faculty of Sciences, Beirut, Lebanon\label{aff124}
\and
Departamento de F\'isica, FCFM, Universidad de Chile, Blanco Encalada 2008, Santiago, Chile\label{aff125}
\and
Institut d'Estudis Espacials de Catalunya (IEEC),  Edifici RDIT, Campus UPC, 08860 Castelldefels, Barcelona, Spain\label{aff126}
\and
Satlantis, University Science Park, Sede Bld 48940, Leioa-Bilbao, Spain\label{aff127}
\and
Institute of Space Sciences (ICE, CSIC), Campus UAB, Carrer de Can Magrans, s/n, 08193 Barcelona, Spain\label{aff128}
\and
Instituto de Astrof\'isica e Ci\^encias do Espa\c{c}o, Faculdade de Ci\^encias, Universidade de Lisboa, Tapada da Ajuda, 1349-018 Lisboa, Portugal\label{aff129}
\and
Cosmic Dawn Center (DAWN)\label{aff130}
\and
Niels Bohr Institute, University of Copenhagen, Jagtvej 128, 2200 Copenhagen, Denmark\label{aff131}
\and
Universidad Polit\'ecnica de Cartagena, Departamento de Electr\'onica y Tecnolog\'ia de Computadoras,  Plaza del Hospital 1, 30202 Cartagena, Spain\label{aff132}
\and
Infrared Processing and Analysis Center, California Institute of Technology, Pasadena, CA 91125, USA\label{aff133}
\and
Dipartimento di Fisica e Scienze della Terra, Universit\`a degli Studi di Ferrara, Via Giuseppe Saragat 1, 44122 Ferrara, Italy\label{aff134}
\and
Istituto Nazionale di Fisica Nucleare, Sezione di Ferrara, Via Giuseppe Saragat 1, 44122 Ferrara, Italy\label{aff135}
\and
INAF, Istituto di Radioastronomia, Via Piero Gobetti 101, 40129 Bologna, Italy\label{aff136}
\and
Instituto de Astrof\'isica de Canarias (IAC); Departamento de Astrof\'isica, Universidad de La Laguna (ULL), 38200, La Laguna, Tenerife, Spain\label{aff137}
\and
Universit\'e PSL, Observatoire de Paris, Sorbonne Universit\'e, CNRS, LERMA, 75014, Paris, France\label{aff138}
\and
Universit\'e Paris-Cit\'e, 5 Rue Thomas Mann, 75013, Paris, France\label{aff139}
\and
Department of Mathematics and Physics E. De Giorgi, University of Salento, Via per Arnesano, CP-I93, 73100, Lecce, Italy\label{aff140}
\and
INFN, Sezione di Lecce, Via per Arnesano, CP-193, 73100, Lecce, Italy\label{aff141}
\and
INAF-Sezione di Lecce, c/o Dipartimento Matematica e Fisica, Via per Arnesano, 73100, Lecce, Italy\label{aff142}
\and
INAF - Osservatorio Astronomico di Brera, via Emilio Bianchi 46, 23807 Merate, Italy\label{aff143}
\and
INAF-Osservatorio Astronomico di Brera, Via Brera 28, 20122 Milano, Italy, and INFN-Sezione di Genova, Via Dodecaneso 33, 16146, Genova, Italy\label{aff144}
\and
ICL, Junia, Universit\'e Catholique de Lille, LITL, 59000 Lille, France\label{aff145}
\and
ICSC - Centro Nazionale di Ricerca in High Performance Computing, Big Data e Quantum Computing, Via Magnanelli 2, Bologna, Italy\label{aff146}
\and
Instituto de F\'isica Te\'orica UAM-CSIC, Campus de Cantoblanco, 28049 Madrid, Spain\label{aff147}
\and
CERCA/ISO, Department of Physics, Case Western Reserve University, 10900 Euclid Avenue, Cleveland, OH 44106, USA\label{aff148}
\and
Laboratoire Univers et Th\'eorie, Observatoire de Paris, Universit\'e PSL, Universit\'e Paris Cit\'e, CNRS, 92190 Meudon, France\label{aff149}
\and
Departamento de F{\'\i}sica Fundamental. Universidad de Salamanca. Plaza de la Merced s/n. 37008 Salamanca, Spain\label{aff150}
\and
Universit\'e de Strasbourg, CNRS, Observatoire astronomique de Strasbourg, UMR 7550, 67000 Strasbourg, France\label{aff151}
\and
Center for Data-Driven Discovery, Kavli IPMU (WPI), UTIAS, The University of Tokyo, Kashiwa, Chiba 277-8583, Japan\label{aff152}
\and
Dipartimento di Fisica - Sezione di Astronomia, Universit\`a di Trieste, Via Tiepolo 11, 34131 Trieste, Italy\label{aff153}
\and
California Institute of Technology, 1200 E California Blvd, Pasadena, CA 91125, USA\label{aff154}
\and
Department of Physics \& Astronomy, University of California Irvine, Irvine CA 92697, USA\label{aff155}
\and
CEA Saclay, DFR/IRFU, Service d'Astrophysique, Bat. 709, 91191 Gif-sur-Yvette, France\label{aff156}
\and
Department of Computer Science, Aalto University, PO Box 15400, Espoo, FI-00 076, Finland\label{aff157}
\and
Instituto de Astrof\'\i sica de Canarias, c/ Via Lactea s/n, La Laguna 38200, Spain. Departamento de Astrof\'\i sica de la Universidad de La Laguna, Avda. Francisco Sanchez, La Laguna, 38200, Spain\label{aff158}
\and
Ruhr University Bochum, Faculty of Physics and Astronomy, Astronomical Institute (AIRUB), German Centre for Cosmological Lensing (GCCL), 44780 Bochum, Germany\label{aff159}
\and
Department of Physics and Astronomy, Vesilinnantie 5, 20014 University of Turku, Finland\label{aff160}
\and
Serco for European Space Agency (ESA), Camino bajo del Castillo, s/n, Urbanizacion Villafranca del Castillo, Villanueva de la Ca\~nada, 28692 Madrid, Spain\label{aff161}
\and
ARC Centre of Excellence for Dark Matter Particle Physics, Melbourne, Australia\label{aff162}
\and
Centre for Astrophysics \& Supercomputing, Swinburne University of Technology,  Hawthorn, Victoria 3122, Australia\label{aff163}
\and
Department of Physics and Astronomy, University of the Western Cape, Bellville, Cape Town, 7535, South Africa\label{aff164}
\and
DAMTP, Centre for Mathematical Sciences, Wilberforce Road, Cambridge CB3 0WA, UK\label{aff165}
\and
Kavli Institute for Cosmology Cambridge, Madingley Road, Cambridge, CB3 0HA, UK\label{aff166}
\and
Department of Astrophysics, University of Zurich, Winterthurerstrasse 190, 8057 Zurich, Switzerland\label{aff167}
\and
Institute for Theoretical Particle Physics and Cosmology (TTK), RWTH Aachen University, 52056 Aachen, Germany\label{aff168}
\and
IRFU, CEA, Universit\'e Paris-Saclay 91191 Gif-sur-Yvette Cedex, France\label{aff169}
\and
Oskar Klein Centre for Cosmoparticle Physics, Department of Physics, Stockholm University, Stockholm, SE-106 91, Sweden\label{aff170}
\and
Astrophysics Group, Blackett Laboratory, Imperial College London, London SW7 2AZ, UK\label{aff171}
\and
Univ. Grenoble Alpes, CNRS, Grenoble INP, LPSC-IN2P3, 53, Avenue des Martyrs, 38000, Grenoble, France\label{aff172}
\and
Dipartimento di Fisica, Sapienza Universit\`a di Roma, Piazzale Aldo Moro 2, 00185 Roma, Italy\label{aff173}
\and
Centro de Astrof\'{\i}sica da Universidade do Porto, Rua das Estrelas, 4150-762 Porto, Portugal\label{aff174}
\and
HE Space for European Space Agency (ESA), Camino bajo del Castillo, s/n, Urbanizacion Villafranca del Castillo, Villanueva de la Ca\~nada, 28692 Madrid, Spain\label{aff175}
\and
Department of Astrophysical Sciences, Peyton Hall, Princeton University, Princeton, NJ 08544, USA\label{aff176}
\and
Theoretical astrophysics, Department of Physics and Astronomy, Uppsala University, Box 515, 751 20 Uppsala, Sweden\label{aff177}
\and
Mathematical Institute, University of Leiden, Einsteinweg 55, 2333 CA Leiden, The Netherlands\label{aff178}
\and
Institute of Astronomy, University of Cambridge, Madingley Road, Cambridge CB3 0HA, UK\label{aff179}
\and
Space physics and astronomy research unit, University of Oulu, Pentti Kaiteran katu 1, FI-90014 Oulu, Finland\label{aff180}
\and
Department of Physics and Astronomy, Lehman College of the CUNY, Bronx, NY 10468, USA\label{aff181}
\and
American Museum of Natural History, Department of Astrophysics, New York, NY 10024, USA\label{aff182}
\and
Center for Computational Astrophysics, Flatiron Institute, 162 5th Avenue, 10010, New York, NY, USA\label{aff183}
\and
Department of Physics and Astronomy, University of British Columbia, Vancouver, BC V6T 1Z1, Canada\label{aff184}}    
%
%
%
%


%
%
\abstract{
The Euclid Wide Survey (EWS) is expected to identify of order $100\,000$ galaxy-galaxy strong lenses across $14\,000\deg^2$. The Euclid Quick Data Release (Q1) of $63.1\deg^2$ \emph{Euclid} images provides an excellent opportunity to test our lens-finding ability, and to verify the anticipated lens frequency in the EWS. Following the Q1 data release, eight machine learning networks from five teams were applied to approximately one million images. This was followed by a citizen science inspection of a subset of around $100\,000$ images, of which $65\%$ received high network scores, with the remainder randomly selected. The top scoring outputs were inspected by experts to establish confident (grade A), likely (grade B), possible (grade C), and unlikely lenses. In this paper we combine the citizen science and machine learning classifiers into an ensemble, demonstrating that a combined approach can produce a purer and more complete sample than the original individual classifiers. Using the expert-graded subset as ground truth, we find that this ensemble can provide a purity of $\rt{52\pm2}\%$ (grade A/B lenses) with $50\%$ completeness (for context, due to the rarity of lenses a random classifier would have a purity of $0.05\%$). We discuss future lessons for the first major \emph{Euclid} data release (DR1), where the big-data challenges will become more significant and will require analysing more than $\sim300$\,million galaxies, and thus time investment of both experts and citizens must be carefully managed. 
}
%
%
%
%
   \titlerunning{\Euclid\/ Q1: The Strong Lensing Discovery Engine -- Ensemble Classification}
   \authorrunning{Euclid Collaboration: P.~Holloway et al.}
   
   \maketitle
%
%
%
%
\section{\label{Introduction}Introduction}
Strong gravitational lensing, whereby light is deflected by gravity to form multiple images of the background source, provides a useful probe for a wide range of science cases, including cosmology (e.g., \citealp{TDCOSMO2023,Li2024}), probing dark matter (e.g., \citealp{Powell2023,WagnerCarena2024}), galaxy evolution \citep{Sonnenfeld2019,Etherington2023}, and probing the early Universe (e.g., \citealp{VanDokkum2024}). With the commencement of the Euclid Wide Survey (EWS, \citealp{Scaramella-EP1}), and the forthcoming start of the Legacy Survey of Space and Time (LSST, \citealp{Ivezic2019}), the field of strong lensing will undergo significant advances, with the number of strong lens systems known expected to increase to around $100\,000$ \citep{Collett2015,Holloway2023,Pearon2024MNRAS}. Furthermore, the large-scale spectroscopic confirmation of perhaps $10\,000$ of these systems using the 4Most Strong Lensing Spectroscopic Legacy Survey (4SLSLS, \citealp{Collett2023}) will allow for population-level analysis on a scale not seen to date.

The \emph{Euclid} satellite \citep{EuclidSkyOverview}, which launched on 1 July 2023, aims to survey $14\,000\deg^2$ of the sky. 
In comparison to LSST, a ground-based seeing-limited (\ang{;;0.8}) survey, the EWS will provide higher-resolution imaging (\ang{;;0.16} in \IE, \citealp{EuclidSkyVIS,Q1-TP002}) over a comparable area, but to a shallower depth (LSST: $r\simeq26.9$\footnote{\url{https://www.lsst.org/scientists/keynumbers}}, \emph{Euclid}: $\IE\simeq26.2$, \citealp{Scaramella-EP1}). The \Euclid Quick Data Release (Q1, \citealp{Q1cite}), comprising of \rt{$63.1\deg^2$} of imaging, provides an excellent test-bed for current strong-lens-detection algorithms. 

To date, strong lenses have been found by a range of different methods and classifiers (see \citealp{Lemon2024} for a review), including arc- and ring-finding algorithms (e.g., \citealp{SeidelBartelmann2007,Sonnenfeld2018}), machine learning (hereafter ML), such as neural network classifiers (e.g., \citealp{PearceCasey2024,More2024,Melo2024,Canameras2024,Nagam2025}), through visual inspection by strong lens experts \citep{Faure2008,Jackson2008,Pawase2014}, and crowd-sourced through citizen science (e.g., \citealp{Marshall2016,More2016,Sonnenfeld2020,Garvin2022,Gonzalez2025}). The citizen science search conducted by \citet{Garvin2022} used archival Hubble Space Telescope (HST) data over roughly half the area ($27\deg^2$) of the Q1 data, identifying 167 A or B-grade candidates. 
Due to the wide range of possible lens configurations, the rarity of strong lens systems, and the high rate of strong-lens mimics (e.g., high-redshift/faint spiral galaxies, ring galaxies, chance alignments, etc.) that vastly outnumber true lenses, state-of-the-art lens classifiers still suffer from a `false positive problem' \citep{Holloway2024}, where non-lenses dominate the high-scoring sample from lens classifiers. To date, these false positives have had to be identified and removed manually.

The EWS presents a challenge to the strong lensing community. The methods used to find the lens systems must be scalable to the billions of galaxy systems this survey will detect. Furthermore, since strong lensing is a rare phenomenon, these methods must either have sufficiently low false positive rates (FPR $\lesssim10^{-3}$) that their lens samples are not dominated by non-lenses, or any subsequent analysis must be done in a manner that accounts for possible contamination (Holloway et al. 2025 in prep.). In \citet{Holloway2024}, the benefits of an ensemble classifier were investigated using data from the Hyper-Suprime Cam (HSC) Subaru Strategic Program \citep{Aihara2019}, whereby multiple lens classifiers (both ML and citizen science) were combined together to provide a purer and more complete lens sample. In this work, we extend this analysis to \emph{Euclid} data, and investigate the scalability of the \emph{Euclid} Strong Lens Discovery Engine to the full EWS. Ensemble classifiers have been used in previous lens searches (e.g., \citealp{Schaefer2018,Andika2023,Canameras2024,Nagam2025}); typically averaging the scores of individual ML classifiers. Here we include an additional calibration step to account for the variation in performance between classifiers. 
This paper aims to answer the following questions.
\begin{enumerate}
    \item How well can such an ensemble of strong lens classifiers perform when applied to high-resolution space-based data?
    \item How does this classifier performance translate into the resulting purity and completeness, and based on this, how would this translate into the number of systems requiring inspection in forthcoming \emph{Euclid} data releases?
    \item Given the large number of strong lens candidates anticipated in DR1 and future releases, is it possible to use inspection by citizens in lieu of using strong lensing experts?
    \item What will the best strategy be for future lens searches to make best use of the individual strengths of a diverse range of strong lens classifiers?
\end{enumerate}
This paper is part of a series outlining the findings of a strong lens search in the Q1 data. \citet[][hereafter Paper A]{Q1-SP048}, details the source selection, search procedure, and final strong lens candidate catalogue. \citet[][Paper B]{Q1-SP052} details lens candidates identified in Q1 data after pre-selecting high-velocity galaxies from the \emph{Sloan} Digital Sky Survey (SDSS) and from Dark Energy Spectroscopic Instrument (DESI) spectra, in addition to the generation of one of the training sets used for the ML and citizen science searches. \citet[][hereafter Paper C]{Q1-SP053} details the primary machine learning models used for finding lenses in this search.  \citet[][Paper D]{Q1-SP054} focusses on the double-source plane lens (DSPL) candidates, including modelling of these systems and the outlook for the full DSPL \emph{Euclid} sample.

This paper is structured as follows. In Sect.\,\ref{Data} we describe the data used in the Q1 strong lens search, the individual classifiers applied to these data, and the expert-grading of lens candidates. In Sect.\,\ref{Classifier_Calibration} we calibrate each of these classifiers, and combine these into an ensemble in Sect.\,\ref{Classifer_Combination}. We discuss results in Sect.\,\ref{Results}, including performance of a range of ensemble classifiers (Sect. \ref{Ensemble_Classifier_Performance}), and the strengths and weaknesses of citizen and ensemble classifiers (Sect. \ref{Strengths_Weaknesses}). We then discuss the outlook for \emph{Euclid} DR1 (Sect.\,\ref{DR1_Outlook}), and conclude in Sect.\,\ref{Conclusion}.

\section{\label{Data}  Data}
For the Q1 strong lens search, we selected \IE-detected extended objects with $\IE<22.5$ which did not have {\it Gaia} counterparts from the Q1 MER catalogue (\citealp{Q1-TP004}, see Paper A for the complete selection). This selection produced $1.09\,\times\,10^6$ sources. Cutout images were then generated using the ESA Science Archive Service and the ESA Datalabs platform \citep{Navarro2024}. In this work, we use \rt{eight} ML networks produced by \rt{five} different teams, applied to these sources, along with classifications for roughly $10$\% of these from the Space Warps citizen science search\footnote{\url{https://www.zooniverse.org/projects/aprajita/space-warps-esa-euclid}} (as described in Paper A). 
For clarity, we briefly describe the individual classifiers used in this ensemble below, but direct readers to Paper A for a comprehensive overview of the source selection, search procedure, and final catalogue of identified Q1 lenses, and Paper C for detailed descriptions of the performance and architectures of the primary \rt{five} networks (Models 1, 2a, 3a, 4, and 5 listed below). In this work we also use 3 additional networks (Models 2b, 3b, and 3c described below) prepared by these teams. Each classifier was shown $10\arcsec$ cutouts. These cutouts were generated using one or both of two colour scalings: `arcsinh' and `MTF' (i.e., `midtone transfer function', defined in Paper A), using different combinations of the \emph{Euclid} band-passes, as described below and summarised in Table \ref{T:Model_Summary}.
\begin{itemize}
    \item Model 0: Space Warps (see Paper A). The Space Warps strong lens search involved around \rt{$1000$} citizens who made a total of \rt{$800\,000$} classifications. Unlike the machine learning classifiers, who were shown all the images in the data set, the citizens classified a subset of \rt{$115\,000$} cutouts, which were either high-scoring objects from the ML classifiers \rt{($80\,000$)}, or randomly drawn from the complete data set \rt{($40\,000$ including overlap)}. Each cutout was classified an average of 7.2 times by citizens; low scoring objects were removed from the platform after six classifications. The citizens were shown both arcsinh and MTF colour settings, using the \IE, \YE, and \JE bands.
    \item Model 1: This network (adapted from \citealp{DominquezSanchez2018,ManjonGarcia}) was a 4-layer Convolutional Neural Network, trained using a range of simulated lens systems, including 64 grade A and B lens candidates from the Galaxy Zoo \citep{Lintott2008} \emph{Euclid} and Cosmic-Dawn projects. The network was applied to the \IE-band-only data set, where the networks' outputs using the MTF and arcsinh colour settings were averaged to produce the final result.
    \item Model 2 (a,b): These OU-100 Convolutional Neural Networks (adapted from \citealp{Wilde2022}) were trained using (a) \IE-band-only, and (b) \IE and \JE bands, respectively, with the MTF colour setting. The training set totalled $32\,000$ non-lenses and simulated lenses, of which $12\%$ were lenses. In addition to simulated lens systems, the training set included around $200$ grade A and B lens candidates identified in \emph{Euclid} imaging. These real \emph{Euclid} lens candidates were identified in searches of Early Release Observation (ERO) data \citep{AcevedoBarroso24}, through inspection of galaxies with spectroscopic data (Paper B), Galaxy Zoo Cosmic Dawn and Galaxy Zoo \emph{Euclid} projects, as well as serendipitous discoveries. Class weights were applied to ensure that non-lenses and lenses were weighted equally overall.
    \item Model 3 (a-c): These networks were adapted from \citet{Leuzzi2024} and had \texttt{IncNet} (a, \citealp{Szegedy2014,Szegedy2015}), \texttt{ResNet} (b, \citealp{He2016,Xie2017}), and \texttt{VGG} (c, \citealp{Simonyan2015}) architectures respectively. They were trained using 40\,000 \IE band-only images (non-lenses and simulated lenses) using the arcsinh colour setting. 
    \item Model 4: \texttt{Zoobot} (see Paper C and \citealp{Walmsley2023}): This Bayesian Neural Network was pre-trained using $9.2\times10^7$ morphological classifications from the Galaxy Zoo project \citep{Lintott2008}, and subsequently fine-tuned using verified non-lenses in Dark Energy Spectroscopic Instrument data (DESI, \citealp{Q1-SP052}) and simulated lenses. This classifier was shown \IE-band-only arcsinh cutouts.
    \item Model 5: This network (derived from \citealp{Chen2020,Oquab2024,Smith2024}) was pre-trained using self-supervised contrastive learning with a Vision Transformer (VT) backbone using 14\,$\times$\,14 patches on $80\,000$ simulated lens images, and $80\,000$ non-lens images, including ring galaxies, mergers, and spirals. The training images used the \IE band and arcsinh scaling. Two augmented views of each image were generated using transformations, including resizing, horizontal flips, colour jitter, and Gaussian blur. A contrastive loss function, scaled by a temperature parameter, was applied to maximise similarity between positive pairs (different views of the same image) and minimise similarity between negative pairs (different images). During fine-tuning, curriculum learning was applied with an initial warm-up phase of five epochs with uniform sampling, followed by dynamic sample weighting and hard-example mining. Hard examples were identified using misclassification or low-confidence predictions, with their weights adjusted progressively to prioritise challenging samples in later epochs.

\end{itemize}
Following the application of the five primary networks (Models 1, 2a, 3a, 4, and 5) and while the Space Warps search was ongoing, the highest scoring $1000$ systems from each of the networks, along with roughly \rt{$2700$} high-scoring systems from Space Warps (with Space Warps score $p_{\rm{SW}}>10^{-5}$) were inspected by strong lensing experts in a phase called `Galaxy Judges' (hereafter GJ, see Paper A). In total, \rt{7700} systems were inspected, identifying \rt{250} grade A systems (confident lens) and \rt{247} grade B systems.
\begin{table*}
\centering
\caption{Summary of the models used in this work, as well as a summary metric of their performance, the purity ($\%$) at 50\% completeness ($P_{50}$), as measured on the test set. The $P_{50}$ value for Model 0 was calculated using the systems in the `random' $40\,000$ data set described in Sect. \ref{Data} for representative comparison. We denote the approximate training set size by $N_{\rm{train}}$ -- in the case of Model 0 (Space Warps), this is given by the median number of training images seen by the citizens.}
\begin{tabular}{cccccc}
\hline
\hline & \\[-1.5ex]
Model & Type & Bands & $N_{\rm{train}}$ & Scaling & $P_{50}$\\
\midrule
0 & CS & \IE, \YE, \JE & 12 & MTF \& arcsinh& 68 \\
1 & CNN (4-Layer)& \IE & $3\times10^4$ & MTF \& arcsinh & 2.1 \\
2a & CNN (OU100) & \IE & $3\times10^4$ & MTF& 0.36 \\
2b & CNN (OU100) & \IE, \JE &$3\times10^4$ & MTF& 0.34 \\
3a & CNN (IncNet) & \IE &$4\times10^4$ & arcsinh& 0.14 \\
3b & CNN (ResNet) & \IE &$4\times10^4$ & arcsinh& 0.08 \\
3c & CNN (VGG) & \IE &$4\times10^4$ &arcsinh& 0.07 \\
4 & CNN (BNN) &\IE &$9\times10^7$& arcsinh& 7.3 \\
5 & VT & \IE &$1.6\times10^5$ & arcsinh& 0.15 \\
\hline
\end{tabular}
\label{T:Model_Summary}
\end{table*}

\section{\label{Method} Method}
\subsection{\label{Classifier_Calibration} Calibration of strong lens classifiers }
To combine multiple classifiers into an ensemble, their outputs need to be calibrated. Here we define calibration as the function that maps the output score of a classifier to the probability that a system with that score is a lens. This calibration accounts for any over- or under-confidence in any particular classifier, perhaps due to differences in training sets, activation functions, or data processing in the images they were shown. The calibration of these classifiers requires a ground truth, a set of objects for which their true type (i.e., lens or non-lens) is known. For this work, since strong lens systems are rare and the majority do not have spectroscopic confirmation, unless stated otherwise, we use the grade A and B lenses from GJ described in Paper A as `true lenses', and assert that all other systems (ungraded or graded as a non-lens) are not lenses. Unless stated explicitly grade C candidates were also treated as non-lenses since in reality these are typically not lensed systems. Therefore, in this work, the probabilities produced following calibration reflect the probability that a system is a grade A or B quality lens. Due to the large number of classifiers applied to this lens search, and because high-scoring systems from all of these were inspected by lens experts and passed to Space Warps, it is reasonable to assume the majority of the lens systems were identified. Analysis in Paper C suggests \rt{$65\%$} of grade A and B lenses were found in the Q1 lens search, with the vast majority of those missing being grade B. The analysis in this paper should therefore be seen as conservative. Since some true lenses will have been labelled here as non-lenses (since non-inspected systems are classed as non-lenses in this work), the model performance will be underestimated. This would likely be most significant at lower model scores, which were not all inspected by experts.

\begin{figure}
\centering
\centering
\includegraphics[width=0.49\textwidth]{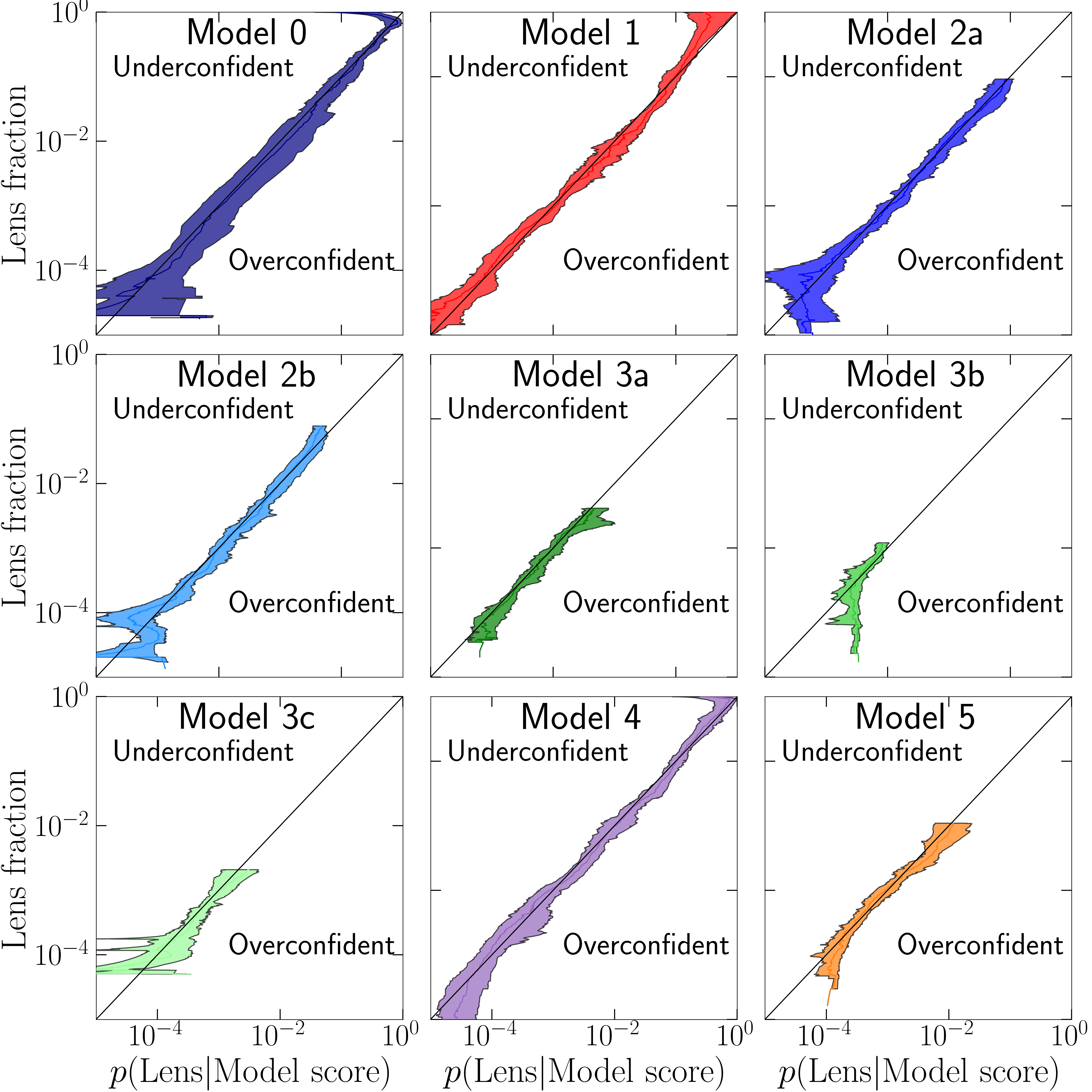}
\caption{Validation of calibration curves, applied to the distinct test set of data. The best performing classifiers can be calibrated up to high probabilities, since their highest-ranked candidates have a high purity. Curves that follow the $y=x$ line are indicative of accurate calibration, where the calibrated probabilities ($x$-axis) match the fraction of grade A+B systems with that score in the test set ($y$-axis). $1\sigma$ uncertainties (shaded regions) are calculated via bootstrapping on the test set.}
\label{Calibration_Validation}
\end{figure}
The best calibration method identified in \citet{Holloway2024} was isotonic regression \citep{Zadrozny2002}. This calibration method assumes that the mapping from the output of the classifier to the lens probability is a monotonically increasing function, i.e., the higher the score that a classifier gives to a particular system, the more likely the system is strongly lensed. 
We split the data into two equal sized data sets, forming a `calibration set' (on which the calibration curves were calculated), and a `test set'. Since we did not adjust the calibration methodology from \citealp{Holloway2024} we did not use a separate validation set. We show the calibration mappings produced by this isotonic regression method in Fig.\,\ref{Calibration_Curves}. We validated the calibration on the test set, as shown in Fig.\,\ref{Calibration_Validation}, where we measure the ratio of the number of lenses to the total number of systems with a given calibrated probability. If the calibration is accurate, this ratio would be equal to the calibrated probability. We find the calibration is accurate across the range of lens classifiers, and across many orders of magnitude. The best performing classifiers (Models 0, 1 and 4) can be calibrated up to high-lens fractions $\mathcal{O}(1)$, while models such as 3b and 3c can only be calibrated over a smaller range, since their highest scoring systems contain a comparatively low fraction of lenses. We find some underconfidence in Model 0 at high lens fractions, indicating some inflexibility in the calibration which assumes the lens fraction increases monotonically with model score. 
Having calibrated each classifier, we combined them into an ensemble, described in Sect.\,\ref{Classifer_Combination}.

\begin{figure}
\centering
\centering
\includegraphics[width=0.45\textwidth]{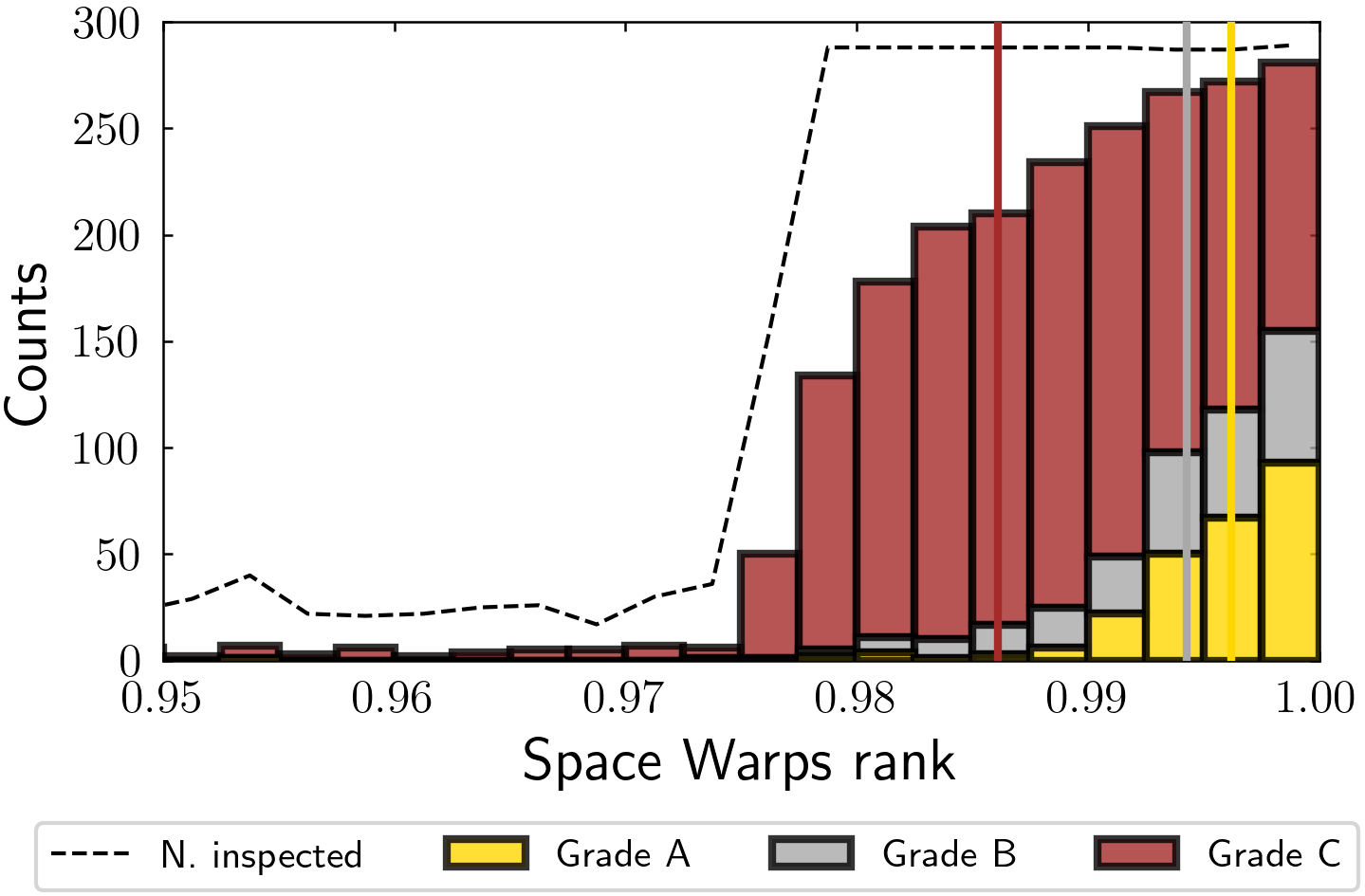}
\caption{GJ grades of the top 5\% ranked subjects from the Space Warps search; the bins are stacked, so the envelope represents the combined purity of grade A+B+C candidates out of the total inspected (dashed line) in each bin. The vertical lines indicate the median rank of lens candidates of each grade; higher-grade lens candidates receive successively higher scores from the citizens.}
\label{SW_Rank_Histogram}
\end{figure}
We also investigate citizen scientist grades as an alternative source of ground truth. Given the number of systems that lensing experts can grade is limited, the possibility of citizens providing equivalent grading is of interest. Figure\,\ref{SW_Rank_Histogram} shows the distribution of grade A, B, and C lens candidates as identified by GJ, in the highest scoring 5\% of systems identified by Space Warps. We find that the citizens are most confident at identifying higher-grade lens systems, which receive successively higher scores. Given this and the strong performance of the citizen scientists (see Model 0 in Fig.\,\ref{Main_ROC_Curve}), we investigated if it was possible to use this classifier as a ground truth for the purposes of calibrating the ML classifiers. While the calibration of Model 0 was overconfident at low lens-fractions, in the primary region of interest (i.e., high probability), the calibration was accurate. We trialled two methods for this to determine if a more rigorous method performed significantly better.
\begin{enumerate}
    \item Calibrating the output scores of Model 0 (Space Warps) using the original GJ ground truth as previously, but then calibrating the remaining models (1--5) using the now-calibrated Model 0 probabilities as a ground truth. To do this,
    for the second calibration of Models 1--5 we drew samples from the calibration set assigning binary classification values according to the calibrated probabilities from Model 0 each time. We then averaged over the calibration curves produced by each set of samples. In this manner, we accounted for the fact the calibration of Model 0 produced probabilities, rather than binary classifications. 
    \item Using a simple threshold in score for Model 0, defining all systems above this threshold as `lens' and vice versa.
\end{enumerate}
We tested both of these methods on the test set, using the original ground truth from GJ and describe the results in Sect. \ref{Results}.

\subsection{\label{Classifer_Combination} Combination of classifiers into an ensemble}
Having calibrated the individual classifiers, they were then combined into an ensemble. The range of classifier types, training data, and network architectures produced outputs that typically show little correlation for the vast majority of the objects in the data set that are not lenses. A non-lens system that ranks relatively highly in one classifier could receive a very low score from a different classifier. This is beneficial, since combining each of these classifiers together can build upon their individual strengths and weaknesses (eight classifiers that were always in complete agreement would provide no more information than one classifier). Following the calibration of each network in Sect.\,\ref{Classifier_Calibration}, we used a Bayesian approach to combine the individual networks, and treated each classifier as effectively independent. From \citet{Holloway2024}, the posterior probability that a given system is strongly lensed (denoted below by $L$), having received calibrated probabilities $\{C_i\}$ from classifiers $\{1,..., N\}$ is given by
\begin{equation}\label{eq:posterior}
    P(L|\{C_i\}) = \frac{N_{\rm L}^{1-N}\prod_N{C_i}}{N_{\rm L}^{1-N}\prod_N{C_i} + N_{\rm NL}^{1-N}\prod_N{(1-C_i)}}\,,
\end{equation}
where we have assumed a prior $p_0 = \frac{N_{\rm L}}{N_{\rm L}+N_{\rm NL}}$, where $N_{\rm L}$ and $N_{\rm NL}$ are the number of lenses and non-lenses in the calibration set (249 and $542\,965$, respectively, in this work). 

The volunteers participating in the Space Warps search were shown a mix of high-scoring lens systems from the ML classifiers, and random systems from the whole data set. Therefore for the majority of objects, only \rt{eight} classifiers were available to form an ensemble. However, scores from all \rt{nine} classifiers ($8\,\times\,$ML + Space Warps) were available for the systems that were most likely be lenses.

\section{\label{Results}Results and Discussion}
\subsection{Ensemble-classifier performance}\label{Ensemble_Classifier_Performance}
\begin{figure*}
\centering
\centering
\includegraphics[width=0.9\textwidth]{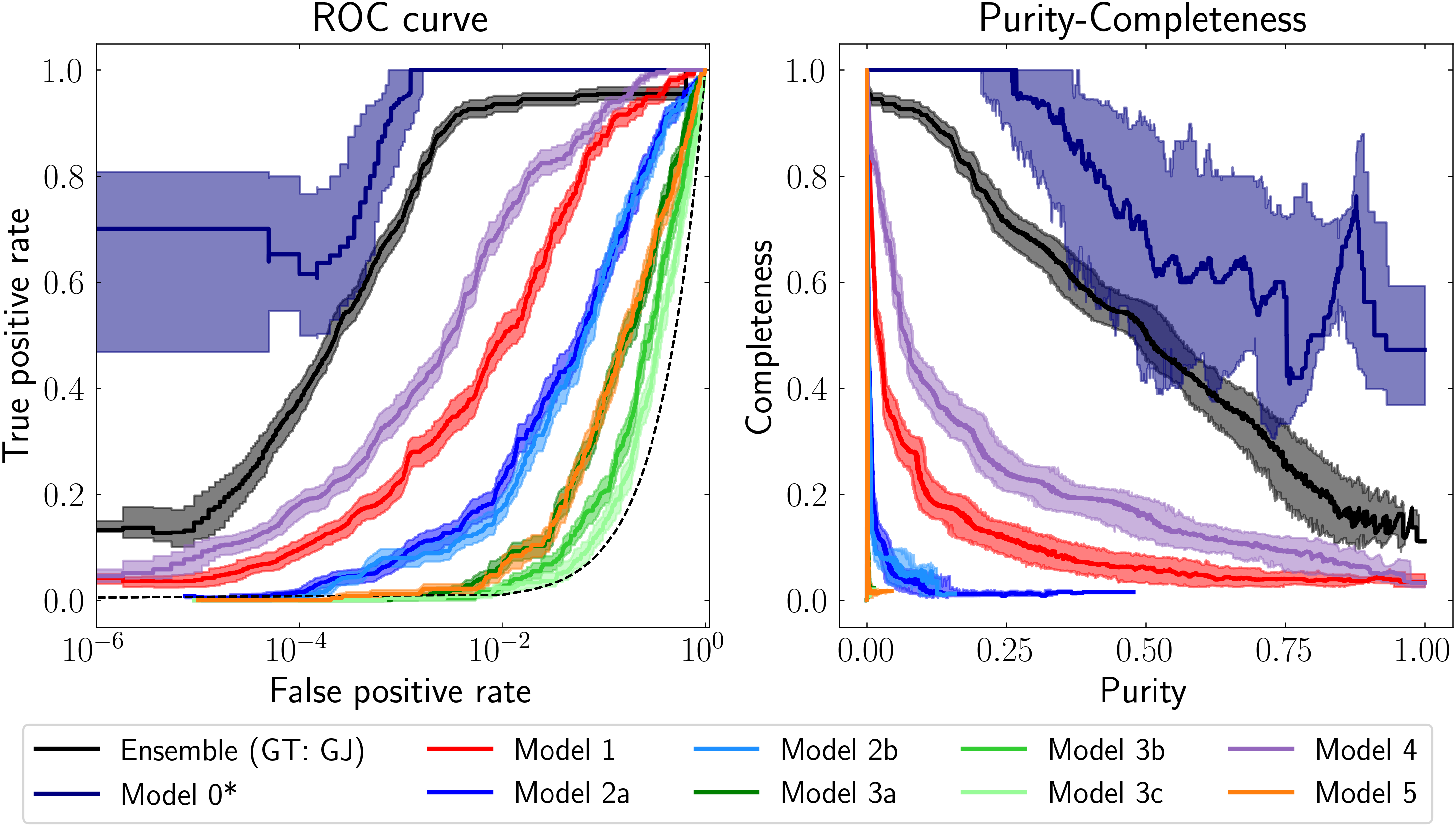}
\caption{ROC (left) and purity-completeness curves (right) of the individual models and an ensemble of all nine models in this work, applied to the test data set. This ensemble was calibrated using the calibration set, assuming a ground truth of GJ grade A and B lenses. $1\sigma$ uncertainties (shaded regions) are calculated via bootstrapping on the test set. The dotted line in the ROC curve indicates the performance of a random classifier.
*Model 0 (Space Warps) was applied to a mix of high-scoring ML systems and randomly selected galaxies. The ROC curve for Model 0 plotted here is generated using only the random subset ($40\,000$) of the full $1$\,million sample, and hence has larger statistical noise. The completeness measurement for this classifier in particular is taken as if the citizens saw the whole data set, which would be unfeasible for future data releases (see Sect.\,\ref{DR1_Outlook}). The curve for the ensemble is generated using the full test set, however.}
\label{Main_ROC_Curve}
\end{figure*}
\begin{figure*}
\centering
\centering
\includegraphics[width=0.9\textwidth]{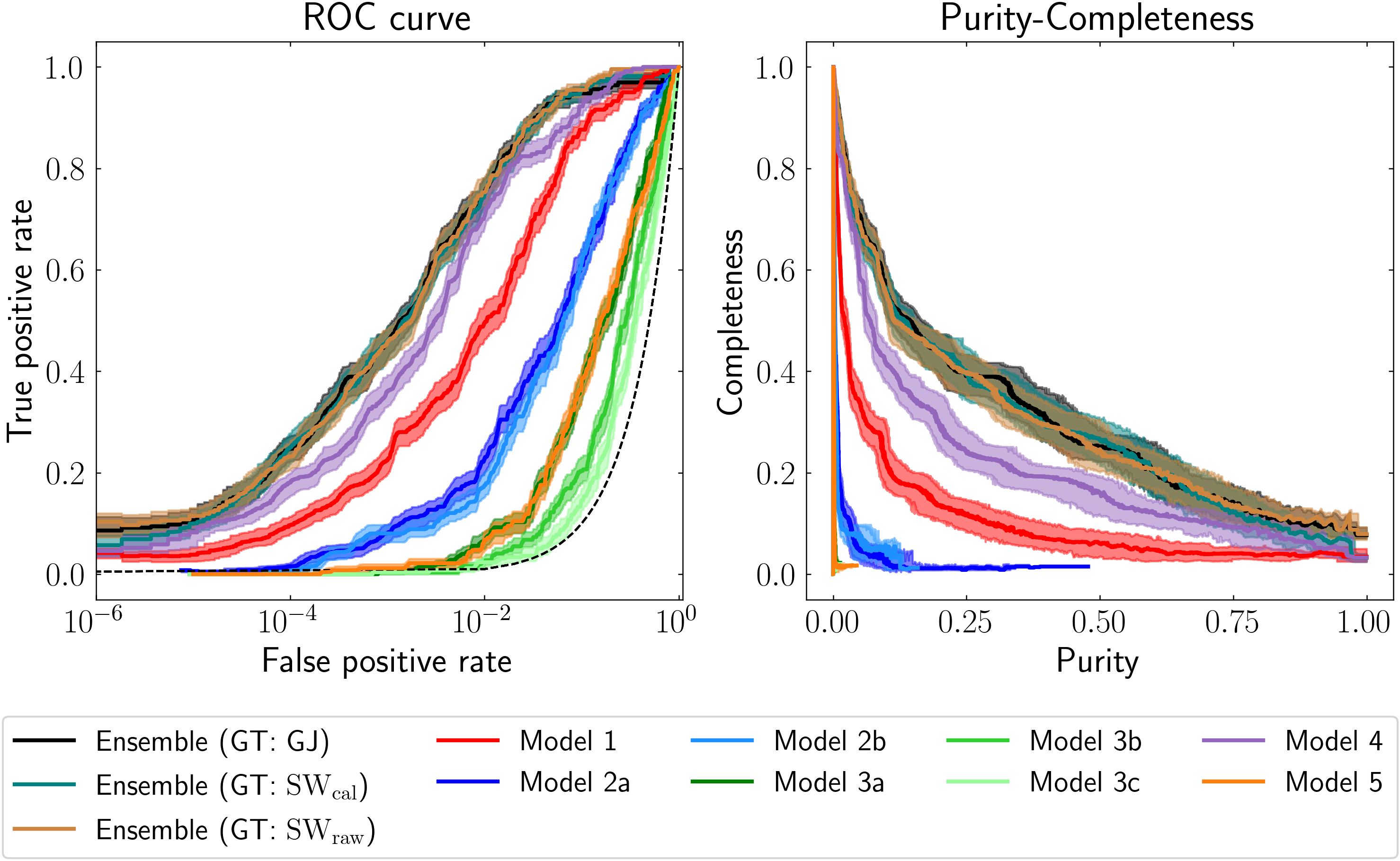}
\caption{ROC (left) and purity-completeness curves (right) of the ML classifiers, and three iterations of an ensemble generated using Model 1, 2a and 4. These ensembles were calibrated using a ground truth of GJ (GT: GJ), Space Warps calibrated probabilities (GT: SW$_{\rm cal}$), and Space Warps binary outputs (GT: SW$_{\rm raw}$, using a raw SW score threshold of $p=1-1.5\times 10^{-8}$)}
\label{NN_Only_ROC_Curve}
\end{figure*}
We find that the ensemble classifier made from all \rt{nine} classifiers provides significant improvement in purity and completeness over the individual ML classifiers. 

Figure\,\ref{Main_ROC_Curve} shows the receiver operating characteristic (ROC) and purity-completeness curves of the individual classifiers, and that of the ensemble. To produce these, we used the expert-grades from GJ as a ground truth (in particular defining grade A and B systems to be lenses, and all other systems, graded or otherwise, to be non-lenses). In this case, the ensemble provides significant improvement over the ML classifiers, achieving 52\% completeness at $50\%$ purity. If grade C systems are excluded entirely from the analysis (i.e., not treated as non-lenses), this metric further improves to 61\% completeness, indicating some remaining high-scoring ensemble systems are grade C candidates. The low purity achieved by some models is indicative of the rarity of strong lens systems; false positive rates $\lesssim10^{-3}$ are required for the resulting sample not to be dominated by non-lenses, which is difficult to achieve. The volunteers participating in the Space Warps search were shown a mix of high-scoring lens systems from the ML classifiers, and random systems from the whole data set. In Fig.\,\ref{Main_ROC_Curve}, we plot the Model 0 results for only the random subset to allow comparison between classifiers. Due to the smaller size of the random data set (and corresponding presence of fewer lenses) there is a much larger statistical noise in this model's curve. The Space Warps classifier is limited in the number of systems it can classify (to about $100\,000$ in this search). 
Figure\,\ref{Main_ROC_Curve} shows that Model 0 performs very well on the random data set achieving \rt{$\gtrsim70\%$} completeness at $50\%$ purity (in the RH panel). However, since lenses are very rare and the total citizen-inspection budget is limited, in the future systems will have to be pre-screened by ML models to maximise the number of lenses identified (see Sect.\,\ref{DR1_Outlook}). The flexibility of the ensemble to account for different numbers of classifiers for each system means it can provide a ranked list of lens candidates across the \emph{whole} data set rather than a subset, and with performance close to that if the citizens had inspected the complete data set.

Figure\,\ref{NN_Only_ROC_Curve} shows the ROC and purity-completeness curves for only the ML classifiers, along with the results from a range of ML-only ensembles. These are generated using the same test set (and with the same GJ ground truth to define the FPR and TPR) as in Fig.\,\ref{Main_ROC_Curve}. Here we focus on the ensemble performance and refer readers to Paper C for greater discussion of the performance of the individual ML models. We find the best ML-only ensembles (which are plotted in Fig.\,\ref{NN_Only_ROC_Curve}) are generated using a subset of the ML classifiers (in particular, Models 1, 2a, and 4). This permutation of models was identified by adding these classifiers one by one to the ensemble until the performance peaked. The difference between an ensemble of Models 1,2a and 4 versus Models 1-5 was marginal and within the uncertainty. Given the very rapid timescale in which the networks had to be trained and optimised for Q1 data (about $1\,$week), the relative performance of the networks used in this work would likely change, and the individual networks would further improve significantly, in future data releases. As described above, we generated these ensembles by calibrating the networks using three different ground truths; the original GJ grades, the Space Warps probabilities ($\rm{SW}_{Cal}$, themselves calibrated using GJ), and the raw Space Warps output ($\rm{SW}_{raw}$, calibrating the networks by defining a `lens' to be all systems with $p\geq1-1.5\times 10^{-8}$, based on the position of the `knee' in the Space Warps ROC curve). We find the benefits of combining the networks to be smaller than that of combining the ML and Space Warps classifiers in Fig.\,\ref{Main_ROC_Curve}. However, we do see an improvement in classification using all of these ground truths for calibration. This hints at the possibility of citizens substituting for expert grading at the larger scales of the forthcoming data releases, which we discuss further in Sect.\,\ref{DR1_Outlook}.
In Appendix \ref{Ap: Effect of Calibation} we demonstrate that calibration is a necessary step prior to combining different models into an ensemble. We find that simply averaging the uncalibrated model outputs produces a much lower performing classifier than first applying calibration and then combining them via the Bayesian framework used in this work.

We also investigated the scatter between the calibrated probabilities produced by the models. We found that this scatter loosely correlated with the error ($|\rm{Truth}-\rm{Pred}|$) on the ensemble probability, i.e., systems were more likely to be misclassified by the ensemble when there was greater disagreement between constituent models. However, many high-grade systems classified correctly by the ensemble also had large scatter in calibrated probability between models. This derived primarily from the varying performance between models -- only the best performing models could be calibrated up to high-probability values (see Fig. \ref{Calibration_Validation}). This resulted in high scatter by default for likely lens candidates, since only the best models could assign probabilities $\mathcal{O}(1)$. 
We found that the highest correlation in model outputs was between classifiers which shared a common training set (models 3a,b,c and models 2a and 2b). We also found correlation between models 1 and 4 (the best performing ML models), with Spearman’s rank correlation $\rho=0.56$. This correlation decreased ($\rho=0.25$) when only considering grade A and B lenses, suggesting they still identified different types of lens.

\subsection{Systems identified by citizens or ensemble}\label{Strengths_Weaknesses}
\begin{figure*}
\centering
\centering
\includegraphics[width=\textwidth]{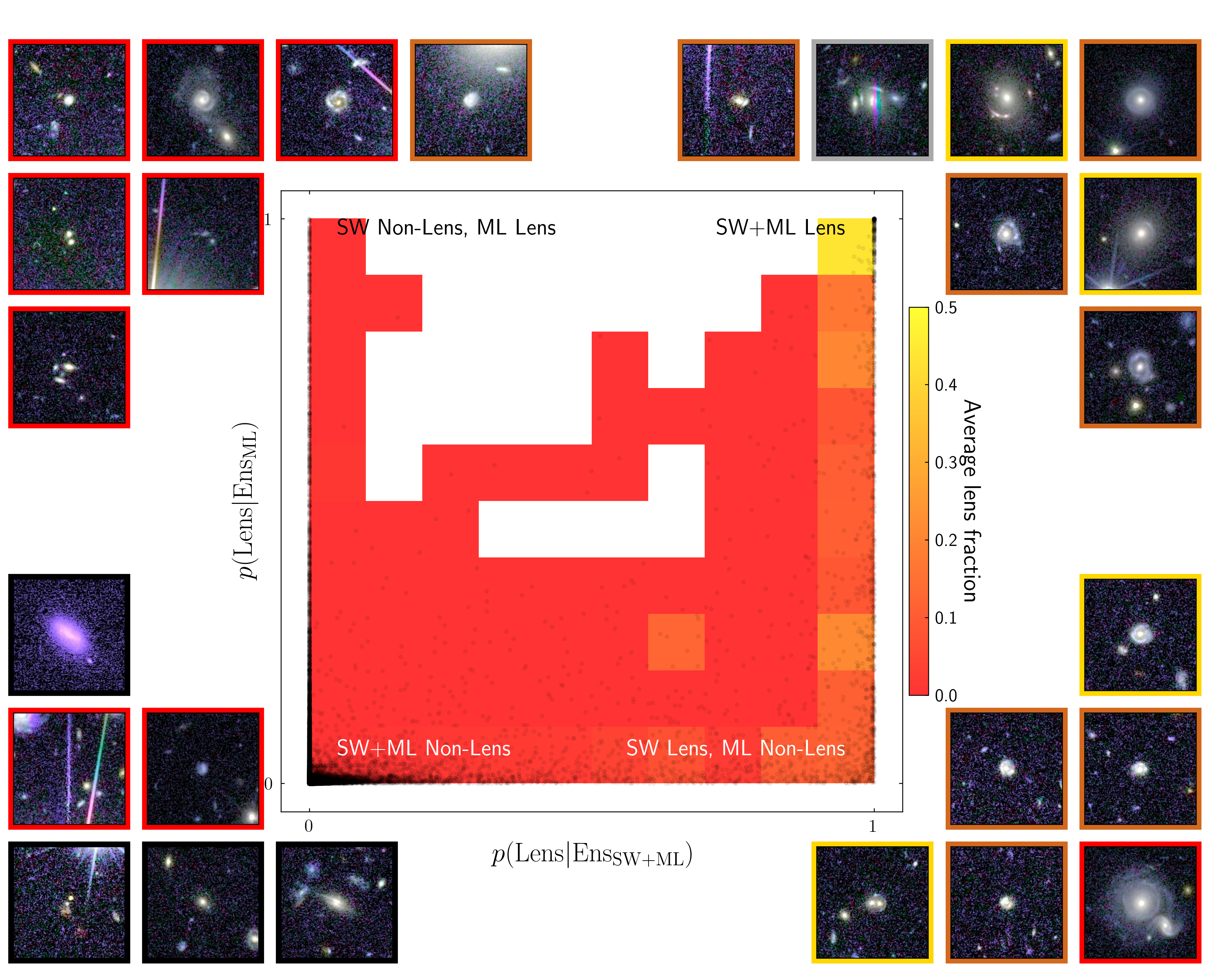}
\caption{Plot of the lens posterior from an ML-only ensemble compared to from a ML+Space Warps ensemble. Systems that receive both a high ML-only posterior and a high ML+Space Warps posterior are the most likely lens candidates. A selection of systems identified by both the ML-only and ML+Space Warps ensembles are shown in the top right, while those only identified by the ML ensemble (ML+Space Warps ensemble) are depicted in the top left (bottom right). Examples of systems rejected by both ensembles are depicted in the bottom left. Cutouts are highlighted by their grade from GJ where available (A: gold, B: silver, C: bronze, Non-lens: red, Ungraded: black). In the central plot, we depict the average (grade A+B) lens fraction, showing that the highest purity can be achieved when both ensembles are in agreement.}
\label{Ensemble_Collage}
\end{figure*}

The ML and citizen science classifiers are naturally very different, and have their own strengths and weaknesses. These may arise from the particular data sets used to train both sets of images (for example, citizens only see a small fraction of the training images that an ML classifier would see), and the intrinsic strengths of the classification method (ML classifiers are excellent at rapid pattern detection, but may struggle with systems which are out-of-distribution such as rare artifacts). Figure\,\ref{Ensemble_Collage} shows the ensemble posterior probability from a ML-only ensemble (from Models 1, 2a, and 4), versus that of a ML + Space Warps ensemble (all \rt{nine classifiers}), along with a selection of cutouts of those identified by the ML-only and ML + Space Warps ensembles. As expected, systems for which both the ML-only and Space Warps + ML ensembles have high posteriors are good lens candidates, including many grade A's. Systems for which the networks produced high scores but which were rejected by citizens have a range of morphologies. These include those with similar arc-like features (such as face-on spiral galaxies), artefacts, and very bright stars that may not have commonly featured in training sets. There were some systems that were ranked highly by citizens but that received lower scores from the networks. Considering just the random sample (i.e., that inspected by both ML and citizen classifiers), \rt{2} (\rt{16}) A/B grade systems were ranked within the top 10\% (1\%) of systems by citizens but did not appear in the top 10\% (1\%) of any network. These include candidates that were mis-centred and those in crowded fields, and are typically B-grade.


Based on the \texttt{PyAutoLens} modelling \citep{Nightingale2021} discussed in Paper A, we investigated correlations in classifier score and modelling properties. Here we restricted our calculations to lens candidates with successful lens models, judged to be lenses (see Paper A). This modelling provided estimates of the lensed/unlensed magnitudes, Einstein radius, and signal-to-noise ratio. We find the ML + Space Warps ensemble classifier score correlates most significantly with magnified source magnitude ($\rho=0.5$), along with total signal-to-noise ratio ($\rho=0.47$). The Space Warps classifier is most correlated with the Einstein radius ($\rho=0.3$), implying the human inspectors are more likely to identify large Einstein radius systems. The selection function of the human inspectors was measured as part of the lens search, and is discussed in Paper A.

\subsection{\label{DR1_Outlook}Outlook for DR1 and future data releases}
\begin{figure}
\centering
\centering
\includegraphics[width=0.45\textwidth]{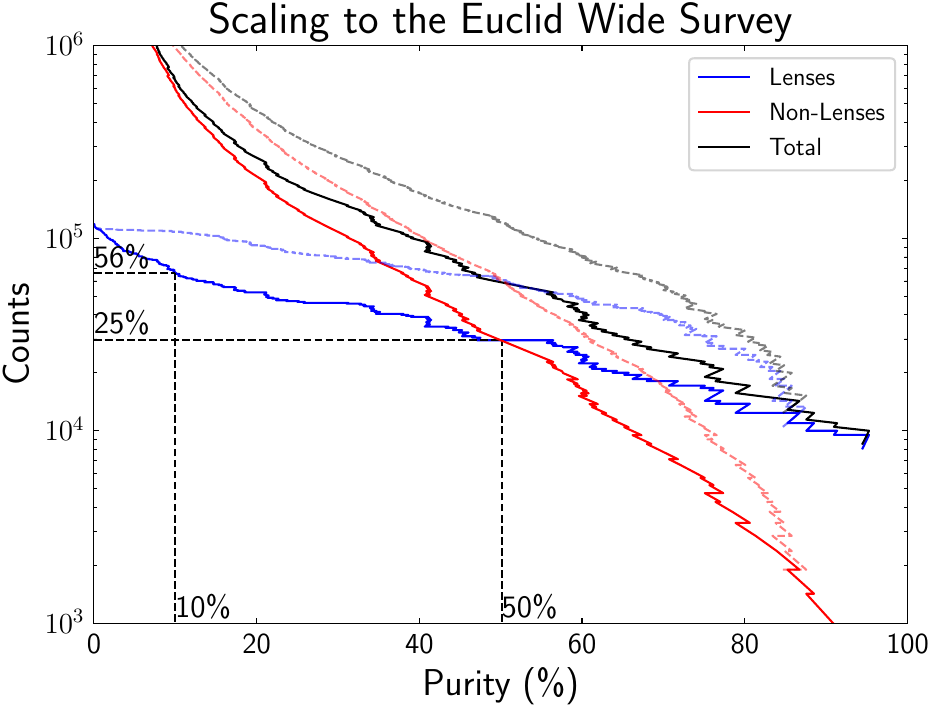}
\caption{Estimate of the number of true and false positives which would be identified by a classifier with the performance of the ensemble classifiers in this work. The bold curves show the performance of the network-only ensemble, while the faint curve shows that of the full \rt{nine-}classifier ensemble; the latter being an optimistic scenario, given citizen classifications likely would not be available for a data set significantly larger than the Q1 data set used in this work. The total number of lens systems has been scaled to reflect the number of grade A and B lenses found in Q1. Completeness values for two illustrative purity thresholds are shown by the dotted lines.}
\label{Scaling_to_ESW}
\end{figure}
In this work, we have used expert grades as a ground truth. With 4SLSLS \citep{Collett2023} it will become possible to use spectroscopically confirmed (and spectroscopically refuted) systems instead, to aid with the calibration of lens classifiers. Prior to this, the ranked outputs of an ensemble classifier in DR1 could also be used to prioritise 4SLSLS follow-up. 
Figure\,\ref{Scaling_to_ESW} shows the number of true lenses and false positives which would be expected as a function of purity in the full EWS, based on the performance of the Space Warps + ML and ML-only ensemble classifiers. For the former ensemble, we find that a small majority of the A/B-grade lenses (\rt{52\%}) would be identified in a 50\% pure sample. Given our definition of `lens' to be grade A or B candidates in this work, it is likely that some of the other 50\% would be grade C candidates. The best ML-only ensemble would achieve \rt{$25\%$} completeness for the same purity, highlighting the value of combining citizen and ML approaches. 

To achieve significantly higher completeness would involve expert inspection of an increasingly large number of false positives, which would rapidly become intractable. However, given the rapid improvement of lens classifiers over time, the much larger catalogue of \emph{Euclid} strong lens systems now available (Paper A) for training and validation, and the fact that the networks applied to Q1 had to be optimised over a very short timescale, it is likely that classification performance will improve further prior to DR1. The large number of classifications of \emph{Euclid} cutouts now available from both citizens and expert graders will provide a rich data set from which to re-train the ML classifiers, particularly on difficult false-positive systems. This will likely occur iteratively following each future \emph{Euclid} data release, allowing the models to continuously improve over time. Similarly, active learning, whereby a ML model is retrained iteratively based on new labels from the most informative systems (see for example \citealp{Walmsley2020,Walmsley2022}) would enable such improvements to occur concurrently with future lens searches. Furthermore, rapid, large-scale modelling of lens systems (e.g., \citealp{Poh2022,Gentile2023,Schuldt2023,Erickson2024}; Busillo et al. in prep.; Venkatraman et al. in prep.) will provide an additional measure for the plausibility of candidate lens systems, and therefore we anticipate that a higher completeness than estimated here is achievable.

\subsubsection{How best to `spend' the visual inspection budgets in DR1}
Unlike ML classifiers, the number of images that both citizens and strong lensing experts can inspect is limited. Therefore, it is crucial to carefully manage what images are shown to citizens to identify the most lenses. For context, in the Q1 Space Warps lens search, around \rt{$1000$} volunteers made \rt{$800\,000$} classifications of \rt{$100\,000$} cutouts over a period of $2$ months. Wider advertisement of the project to the public could boost these numbers. For example, a Space Warps strong lens search using HSC data \citep{Sonnenfeld2020} was featured on a US national radio channel, and received $2.5\times10^6$ classifications from $10\,000$ volunteers, also over a $2$ month period. Furthermore, more relaxed time constraints for DR1 inspection and analysis compared to Q1 will help increase both the citizen and expert inspection budgets. However, there remains a limit to the number that can be inspected. The citizens in the Q1 lens search were shown a mix of cutouts of high-scoring systems, and random cutouts from Q1 data. While showing a small number of randomly selected galaxies can help identify unusual lens configurations, a purely random selection of systems to the volunteers would not include the vast majority of lens systems (only about \rt{$70$} lenses in a random selection of $100\,000$ cutouts using the Q1 pre-selection). This therefore supports a two-stage approach, whereby ML classifiers are applied to the full DR1 data set, of which a subset (likely a few thousand) are inspected by experts to verify the performance of the networks. The ML classifiers would then be collated into an ensemble, and the resulting ranked list could be used to inform which cutouts are shown to citizens.
The strong lensing experts who took part in GJ inspected and graded $7000$ images over roughly $3$ months. The DR1 data set will be $36$ times larger than the Q1 data release, but the inspection capacity of expert graders will likely remain similar. Given the longer timescale, it is possible that experts will inspect a larger sample so we consider two scenarios here. We first scale our results by a factor of $36\times$ to that of the DR1 area (in particular on the number of lenses/non-lenses above a given model threshold), then make cuts based on realistic inspection limits to calculate the total number of lenses which may be found. Based on the test set used in this work (and assuming for the moment that a similar ensemble was produced, perhaps via the calibrated Q1 networks), the highest-ranked $100\,000$ systems from an ML-only ensemble would contain \rt{$9100$} grade A/B lenses, out of a total of \rt{$15\,000$} detectable systems in DR1. By comparison, \rt{$7300$} would be identified using the same cut on Model 1 alone. If the former were shown to citizens, the highest ranked $5000$ from a ML + Space Warps ensemble would contain around \rt{$3900$} grade A/B systems (i.e., a fairly pure but incomplete sample). A simple cut on Space Warps score would produce \rt{$3500$} systems in the top $5000$. In an optimistic scenario, in which the citizens inspect $1\,000\,000$ systems from an ML ensemble, of which the highest-scoring $15\,000$ are passed to experts, we find that approximately \rt{$7600$} would be identified if the scores from all classifiers were first combined into a citizen+ML ensemble. Combining the scores into an ensemble at each stage would therefore be worthwhile to produce a higher purity sample.  

Without significantly increasing the inspection capacity of either experts or citizens, the resulting completeness would be relatively low ($<50\%$). A priority for DR1 will be obtaining a sufficient dataset (around $10\,000$ systems) of high-grade strong lens candidates for followup through 4SLSLS, which based on the current performance of the lens classifiers and anticipated improvements will likely be met. Beyond this, the importance of purity versus completeness will vary depending on the particular science case. While typical lens studies have focussed on spectroscopically confirmed systems, if the contamination rate is known, unbiased inference can be performed on impure data sets (e.g., \citealp{Kunz2007,Roberts2017}; Holloway et al. in prep.). The completeness could be improved through the automated modelling of candidates, or by adding a refinement stage to citizen inspection as done in \citet{Marshall2016}. The latter of these could also involve the citizens grading systems in line with typical lens searches (A/B/C/X), and calibrating their classifications to match those of experts. Additionally, applying the SWAP methodology \citep{Marshall2016} to the final inspection, whereby the final grade was a weighted average of the experts' grades based on their performance on a chosen training set, would increase the efficiency (and thus reduce the time burden) of the expert grading. 
\section{\label{Conclusion}  Summary and conclusions}
In this work, we have produced an ensemble strong lens classifier using the Q1 data release. In answer to the questions set out in Sect.\,\ref{Introduction}, we summarise our conclusions below.
\begin{enumerate}
    \item 
    An ensemble classifier provides significant improvement in classification when both ML and citizen science classifiers are used in the ensemble. In particular, the ensemble can still be used across the whole data set, providing posterior probabilities that each system is a lens, even when some classification data are incomplete (for example where citizens are only shown a subset of the data). 
    \item 
    An ensemble composed of neural networks and citizen scientists produces a \rt{52\%} complete sample at 50\% purity, or \rt{91\%} complete at 10\% purity. An ensemble comprised of only ML classifiers produced a \rt{$25\%$} complete sample, with $50\%$ purity, or \rt{$56\%$} complete at 10\% purity. Due to limited inspection budgets, it is likely that future expert inspected samples will have much higher purity than at present (e.g., $6.8\%$ in this work). 
    \item 
    Citizen classification can produce a high-purity sample of lens candidates, and higher-grade lenses receive progressively higher scores from citizens. Citizens could stand in for expert graders in future searches although care will need to be taken to account for the possibility of misclassifications and the total citizen-inspection budget.
    \item 
    Showing citizens a random selection of cutouts would only result in a small fraction of lens systems being identified, since the vast majority of cutouts would not contain a lens system, and the number of images citizens can inspect is limited. Using a two-stage approach via an ML-only ensemble, whereby citizens are only shown highly-scored systems from this ensemble would significantly increase the total number of lenses identified.
    \item Fine-tuning machine learning classifiers by using ensemble scores from this Q1 search as labels within their training sets would likely diversify the range of lenses that these automated methods could identify. Furthermore, given anticipated lens search campaigns with future data releases, such fine-tuning could be done iteratively as more lenses and non-lenses are classified.
\end{enumerate}
With anticipated improvements in lens classification following this lens search, we expect more than $10\,000$ A/B grade lenses to be identified in \emph{Euclid} DR1, heralding the start of a new era for strong lens science.
%
%

\begin{acknowledgements}
\AckQone
\AckDatalabs
\AckEC  
C.T. acknowledges the INAF grant 2022 LEMON. LL thanks the support from INAF theory Grant 2022: Illuminating Dark Matter using Weak Lensing by Cluster Satellites.

SS has received funding from the European Union’s Horizon 2022 research and innovation programme under the Marie Skłodowska-Curie grant agreement No 101105167 — FASTIDIoUS.

The Dunlap Institute is funded through an endowment established by the David Dunlap family and the University of Toronto.

GD acknowledges the funding by the European Union - NextGenerationEU, in the framework of the HPC project – “National Centre for HPC, Big Data and Quantum Computing” (PNRR - M4C2 - I1.4 - CN00000013 – CUP J33C22001170001).

\end{acknowledgements}

%
%

\bibliography{References,Q1_Bibliography,Euclid}

%

\begin{appendix}
  \onecolumn 
\section{Calibration Curves for each Model}
\begin{figure}
\centering
\centering
\includegraphics[width=\textwidth]{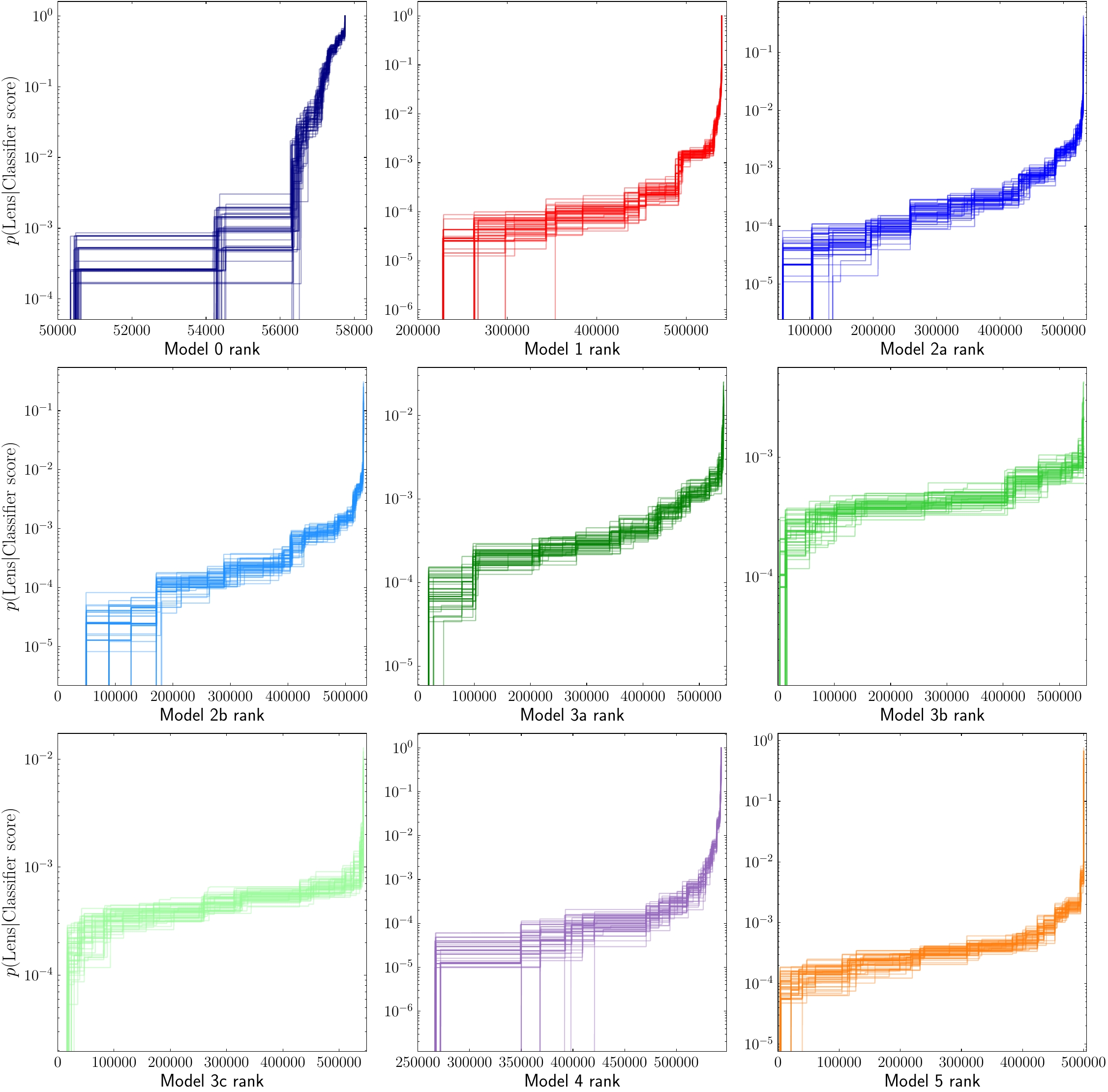}
\caption{Mapping from the ranking of each object in the test set to the calibrated probability, based on the isotonic regression \citep{Zadrozny2002} procedure. We show 50 calibration curves for each model, generated by bootstrapping. The lower $x$-axis limit is trimmed for clarity to where the mapping is non-zero. The limits for Model 0 are more restricted, since the citizens were only shown a subset of the whole data set, and a high proportion of the lowest scoring systems received a calibrated probability of 0.}
\label{Calibration_Curves}
\end{figure}

The mapping from model rank to calibrated probability for each model is shown in Figure \ref{Calibration_Curves}. To apply these calibration mappings to the test set, we interpolate these rank values to the original model scores, such that the calibration function can map between model score and calibrated probability on new data. The best performing models can achieve calibrated probabilities $\mathcal{O}(1)$, indicating their highest scoring systems have a high lens purity.

\section{Effect of Calibration on Classifier Combination}\label{Ap: Effect of Calibation}
\begin{figure*}
\centering
\centering
\includegraphics[width=0.9\textwidth]{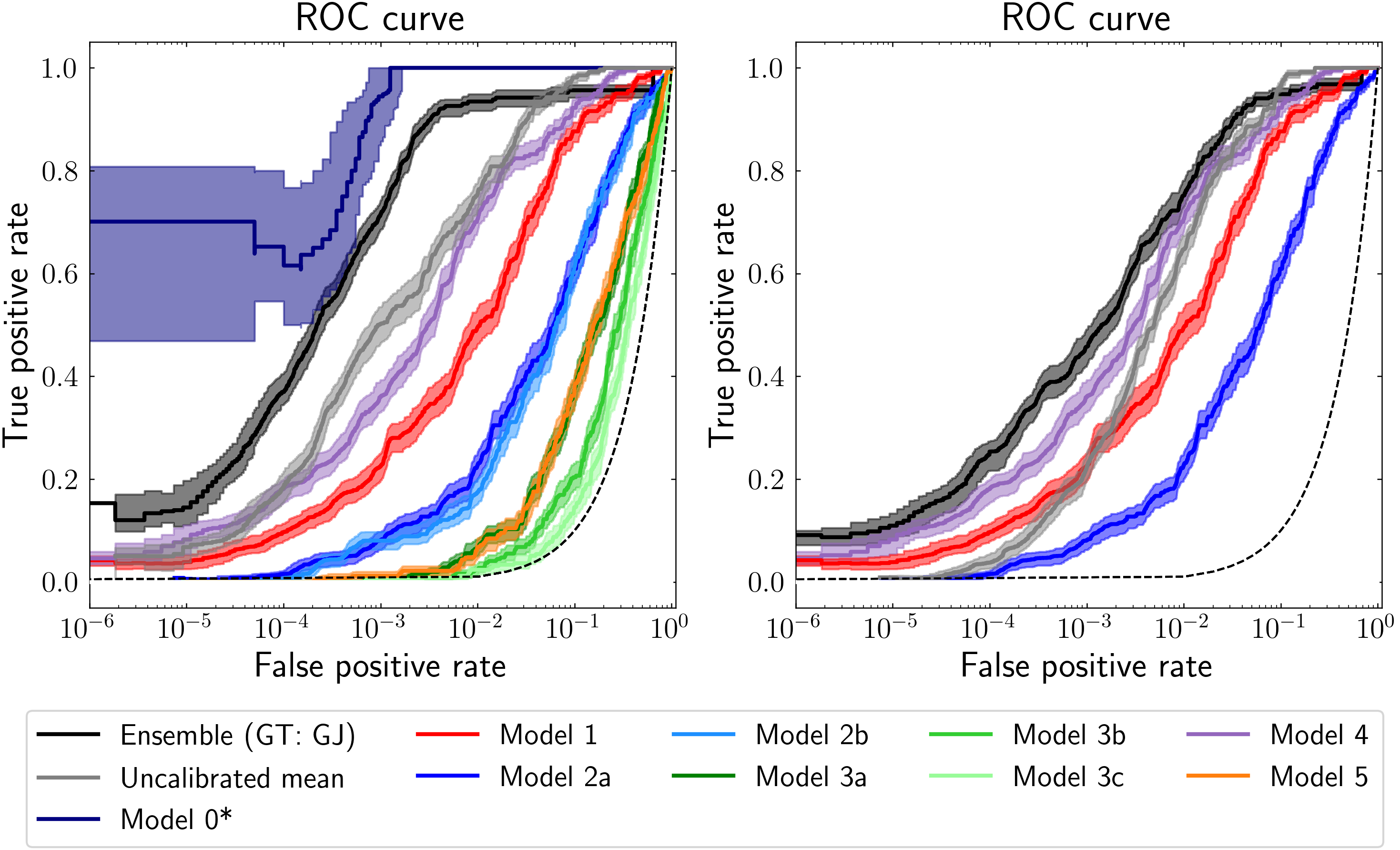}
\caption{Comparison of the ensemble ROC curves generated via calculating the mean average of the uncalibrated model outputs (grey), versus Bayesian combination following calibration (black). The ensemble of Models 1, 2a and 4 is shown on the right, and the ensemble of all models is shown on the left. The ROC curves of the individual classifiers making up these ensembles are also shown as previously.}
\label{Simple_Av_ROC_Curve}
\end{figure*}

Figure \ref{Simple_Av_ROC_Curve} shows the ROC curves generated from simply averaging the uncalibrated model outputs compared to first calibrating each model before combining them in a Bayesian manner. The calibration provides significant improvement over the uncalibrated ensembles. Furthermore the uncalibrated classifier performs much worse than its constituent parts in the case of the ML-only ensemble.

%
\end{appendix}

\end{document}